\documentclass[prl,groupedaddress,superscriptaddress,twocolumn]{revtex4-1}

\usepackage[pdftex]{graphicx}
\usepackage{dcolumn}
\usepackage{bm}
\usepackage{amsmath}
\usepackage{latexsym}
\usepackage{epstopdf}
\usepackage{color}
\usepackage{lipsum}
\usepackage{xargs}
\usepackage{verbatim}

\usepackage{lipsum}

\usepackage{hyperref}
\hypersetup{colorlinks=true, linkcolor=blue, citecolor=blue, urlcolor=blue }

\newcommand{\rfig}[1]{Fig.~\ref{#1}}
\newcommand{\rfigs}[1]{Figs.~\ref{#1}}

\makeindex

\begin{document}
	
	\title{Interaction-induced topological pumping in a solid-state quantum system}
	
	\author{Ziyu Tao}
	\altaffiliation{These authors contributed equally to this work.}
	\affiliation{Shenzhen Institute for Quantum Science and Engineering, Southern University of Science and Technology, Shenzhen 518055, China}
	\affiliation{International Quantum Academy, Shenzhen 518048, China}
	\affiliation{Guangdong Provincial Key Laboratory of Quantum Science and Engineering, Southern University of Science and Technology, Shenzhen 518055, China}

	\author{Wenhui Huang}
	\altaffiliation{These authors contributed equally to this work.}
	\affiliation{Shenzhen Institute for Quantum Science and Engineering, Southern University of Science and Technology, Shenzhen 518055, China}
	\affiliation{International Quantum Academy, Shenzhen 518048, China}
	\affiliation{Guangdong Provincial Key Laboratory of Quantum Science and Engineering, Southern University of Science and Technology, Shenzhen 518055, China}
	
	\author{Jingjing Niu}
	\altaffiliation{These authors contributed equally to this work.}
	\affiliation{International Quantum Academy, Shenzhen 518048, China}
	
	\author{Libo Zhang}
	\altaffiliation{These authors contributed equally to this work.}
	\affiliation{Shenzhen Institute for Quantum Science and Engineering, Southern University of Science and Technology, Shenzhen 518055, China}
	\affiliation{International Quantum Academy, Shenzhen 518048, China}
	\affiliation{Guangdong Provincial Key Laboratory of Quantum Science and Engineering, Southern University of Science and Technology, Shenzhen 518055, China}
		
	\author{Yongguan Ke}
	\affiliation{Guangdong Provincial Key Laboratory of Quantum Metrology and Sensing and School of Physics and Astronomy, Sun Yat-Sen University, Zhuhai 519082, China}
	\affiliation{State Key Laboratory of Optoelectronic Materials and Technologies, Sun Yat-Sen University, Guangzhou 510275, China}
	
	\author{Xiu Gu}
	\affiliation{Shenzhen Institute for Quantum Science and Engineering, Southern University of Science and Technology, Shenzhen 518055, China}
	\affiliation{International Quantum Academy, Shenzhen 518048, China}
	\affiliation{Guangdong Provincial Key Laboratory of Quantum Science and Engineering, Southern University of Science and Technology, Shenzhen 518055, China}

	\author{Ling Lin}
	\affiliation{College of Physics and Optoelectronic Engineering, Shenzhen University, Shenzhen 518060, China}
	 
	\author{Jiawei Qiu}
	\affiliation{Shenzhen Institute for Quantum Science and Engineering, Southern University of Science and Technology, Shenzhen 518055, China}
	\affiliation{International Quantum Academy, Shenzhen 518048, China}
	\affiliation{Guangdong Provincial Key Laboratory of Quantum Science and Engineering, Southern University of Science and Technology, Shenzhen 518055, China}
		 
	\author{Xuandong Sun}
	\affiliation{Shenzhen Institute for Quantum Science and Engineering, Southern University of Science and Technology, Shenzhen 518055, China}
	\affiliation{International Quantum Academy, Shenzhen 518048, China}
	\affiliation{Guangdong Provincial Key Laboratory of Quantum Science and Engineering, Southern University of Science and Technology, Shenzhen 518055, China}
			 
	\author{Xiaohan Yang}
	\affiliation{Shenzhen Institute for Quantum Science and Engineering, Southern University of Science and Technology, Shenzhen 518055, China}
	\affiliation{International Quantum Academy, Shenzhen 518048, China}
	\affiliation{Guangdong Provincial Key Laboratory of Quantum Science and Engineering, Southern University of Science and Technology, Shenzhen 518055, China}

	\author{Jiajian Zhang}
	\affiliation{Shenzhen Institute for Quantum Science and Engineering, Southern University of Science and Technology, Shenzhen 518055, China}
	\affiliation{International Quantum Academy, Shenzhen 518048, China}
	\affiliation{Guangdong Provincial Key Laboratory of Quantum Science and Engineering, Southern University of Science and Technology, Shenzhen 518055, China}
				 
	\author{Jiawei Zhang}
	\affiliation{Shenzhen Institute for Quantum Science and Engineering, Southern University of Science and Technology, Shenzhen 518055, China}
	\affiliation{International Quantum Academy, Shenzhen 518048, China}
	\affiliation{Guangdong Provincial Key Laboratory of Quantum Science and Engineering, Southern University of Science and Technology, Shenzhen 518055, China}

	\author{Shuxiang Zhao}
	\affiliation{Shenzhen Institute for Quantum Science and Engineering, Southern University of Science and Technology, Shenzhen 518055, China}
	\affiliation{International Quantum Academy, Shenzhen 518048, China}
	\affiliation{Guangdong Provincial Key Laboratory of Quantum Science and Engineering, Southern University of Science and Technology, Shenzhen 518055, China}

	\author{Yuxuan Zhou}
	\affiliation{Shenzhen Institute for Quantum Science and Engineering, Southern University of Science and Technology, Shenzhen 518055, China}
	\affiliation{International Quantum Academy, Shenzhen 518048, China}
	\affiliation{Guangdong Provincial Key Laboratory of Quantum Science and Engineering, Southern University of Science and Technology, Shenzhen 518055, China}

	\author{Xiaowei Deng}
	\affiliation{Shenzhen Institute for Quantum Science and Engineering, Southern University of Science and Technology, Shenzhen 518055, China}
	\affiliation{International Quantum Academy, Shenzhen 518048, China}
	\affiliation{Guangdong Provincial Key Laboratory of Quantum Science and Engineering, Southern University of Science and Technology, Shenzhen 518055, China}

	\author{Changkang Hu}
	\affiliation{Shenzhen Institute for Quantum Science and Engineering, Southern University of Science and Technology, Shenzhen 518055, China}
	\affiliation{International Quantum Academy, Shenzhen 518048, China}
	\affiliation{Guangdong Provincial Key Laboratory of Quantum Science and Engineering, Southern University of Science and Technology, Shenzhen 518055, China}

	\author{Ling Hu}
	\affiliation{Shenzhen Institute for Quantum Science and Engineering, Southern University of Science and Technology, Shenzhen 518055, China}
	\affiliation{International Quantum Academy, Shenzhen 518048, China}
	\affiliation{Guangdong Provincial Key Laboratory of Quantum Science and Engineering, Southern University of Science and Technology, Shenzhen 518055, China}

	\author{Jian Li}
	\affiliation{Shenzhen Institute for Quantum Science and Engineering, Southern University of Science and Technology, Shenzhen 518055, China}
	\affiliation{International Quantum Academy, Shenzhen 518048, China}
	\affiliation{Guangdong Provincial Key Laboratory of Quantum Science and Engineering, Southern University of Science and Technology, Shenzhen 518055, China}

	\author{Yang Liu}
	\affiliation{Shenzhen Institute for Quantum Science and Engineering, Southern University of Science and Technology, Shenzhen 518055, China}
	\affiliation{International Quantum Academy, Shenzhen 518048, China}
	\affiliation{Guangdong Provincial Key Laboratory of Quantum Science and Engineering, Southern University of Science and Technology, Shenzhen 518055, China}

	\author{Dian Tan}
	\affiliation{Shenzhen Institute for Quantum Science and Engineering, Southern University of Science and Technology, Shenzhen 518055, China}
	\affiliation{International Quantum Academy, Shenzhen 518048, China}
	\affiliation{Guangdong Provincial Key Laboratory of Quantum Science and Engineering, Southern University of Science and Technology, Shenzhen 518055, China}
	
	\author{Yuan Xu}
	\affiliation{Shenzhen Institute for Quantum Science and Engineering, Southern University of Science and Technology, Shenzhen 518055, China}
	\affiliation{International Quantum Academy, Shenzhen 518048, China}
	\affiliation{Guangdong Provincial Key Laboratory of Quantum Science and Engineering, Southern University of Science and Technology, Shenzhen 518055, China}
	
	\author{Tongxing Yan}
	\affiliation{Shenzhen Institute for Quantum Science and Engineering, Southern University of Science and Technology, Shenzhen 518055, China}
	\affiliation{International Quantum Academy, Shenzhen 518048, China}
	\affiliation{Guangdong Provincial Key Laboratory of Quantum Science and Engineering, Southern University of Science and Technology, Shenzhen 518055, China}

	\author{Yuanzhen Chen} \email{chenyz@sustech.edu.cn}
	\affiliation{Shenzhen Institute for Quantum Science and Engineering, Southern University of Science and Technology, Shenzhen 518055, China}
	\affiliation{International Quantum Academy, Shenzhen 518048, China}
	\affiliation{Guangdong Provincial Key Laboratory of Quantum Science and Engineering, Southern University of Science and Technology, Shenzhen 518055, China}
	\affiliation{Department of Physics, Southern University of Science and Technology, Shenzhen 518055, China} 

	\author{Chaohong Lee} \email{chleecn@szu.edu.cn}
	\affiliation{College of Physics and Optoelectronic Engineering, Shenzhen University, Shenzhen 518060, China}

	\author{Youpeng Zhong}
	\email{zhongyp@sustech.edu.cn}
	\affiliation{Shenzhen Institute for Quantum Science and Engineering, Southern University of Science and Technology, Shenzhen 518055, China}
	\affiliation{International Quantum Academy, Shenzhen 518048, China}
	\affiliation{Guangdong Provincial Key Laboratory of Quantum Science and Engineering, Southern University of Science and Technology, Shenzhen 518055, China}		
	
	\author{Song Liu}
	\affiliation{Shenzhen Institute for Quantum Science and Engineering, Southern University of Science and Technology, Shenzhen 518055, China}
	\affiliation{International Quantum Academy, Shenzhen 518048, China}
	\affiliation{Guangdong Provincial Key Laboratory of Quantum Science and Engineering, Southern University of Science and Technology, Shenzhen 518055, China}		
	
	\author{Dapeng Yu}\email{yudp@sustech.edu.cn}
	\affiliation{Shenzhen Institute for Quantum Science and Engineering, Southern University of Science and Technology, Shenzhen 518055, China}
	\affiliation{International Quantum Academy, Shenzhen 518048, China}
	\affiliation{Guangdong Provincial Key Laboratory of Quantum Science and Engineering, Southern University of Science and Technology, Shenzhen 518055, China}
	\affiliation{Department of Physics, Southern University of Science and Technology, Shenzhen 518055, China} 
		

	\begin{abstract}
	As the basis for generating multi-particle quantum correlations, inter-particle interaction plays a crucial role in collective quantum phenomena, quantum phase transitions, and quantum information processing.
	It can profoundly alter the band structure of quantum many-body systems and give rise to exotic topological phenomena.
	Conventional topological pumping, which has been well demonstrated in driven linear or noninteracting systems, may break down in the presence of strong interaction.
	However, the interplay between band topology and interaction could also induce emergent topological pumping of interacting particles, but its experimental realization has proven challenging.
	Here we demonstrate interaction-induced topological pumping in a solid-state quantum system comprising an array of 36 superconducting qubits.
	With strong interaction inherent in the qubits and site-resolved controllability of the lattice potential and hopping strength, we realize the topological Thouless pumping of single and two bounded particles.
	Beyond these topological phenomena with linear or noninteracting counterparts, we also observe topologically resonant tunneling and asymmetric edge-state transport of interacting particles.
	Our work creates a paradigm for multi-particle topological effects, and provides a new pathway to the study of exotic topological phenomena, many-body quantum transport, and quantum information transfer.
	\end{abstract}
		
	\maketitle	
	
	
	\begin{figure*}[t]
		\begin{center}
			\includegraphics[width=0.9\textwidth]{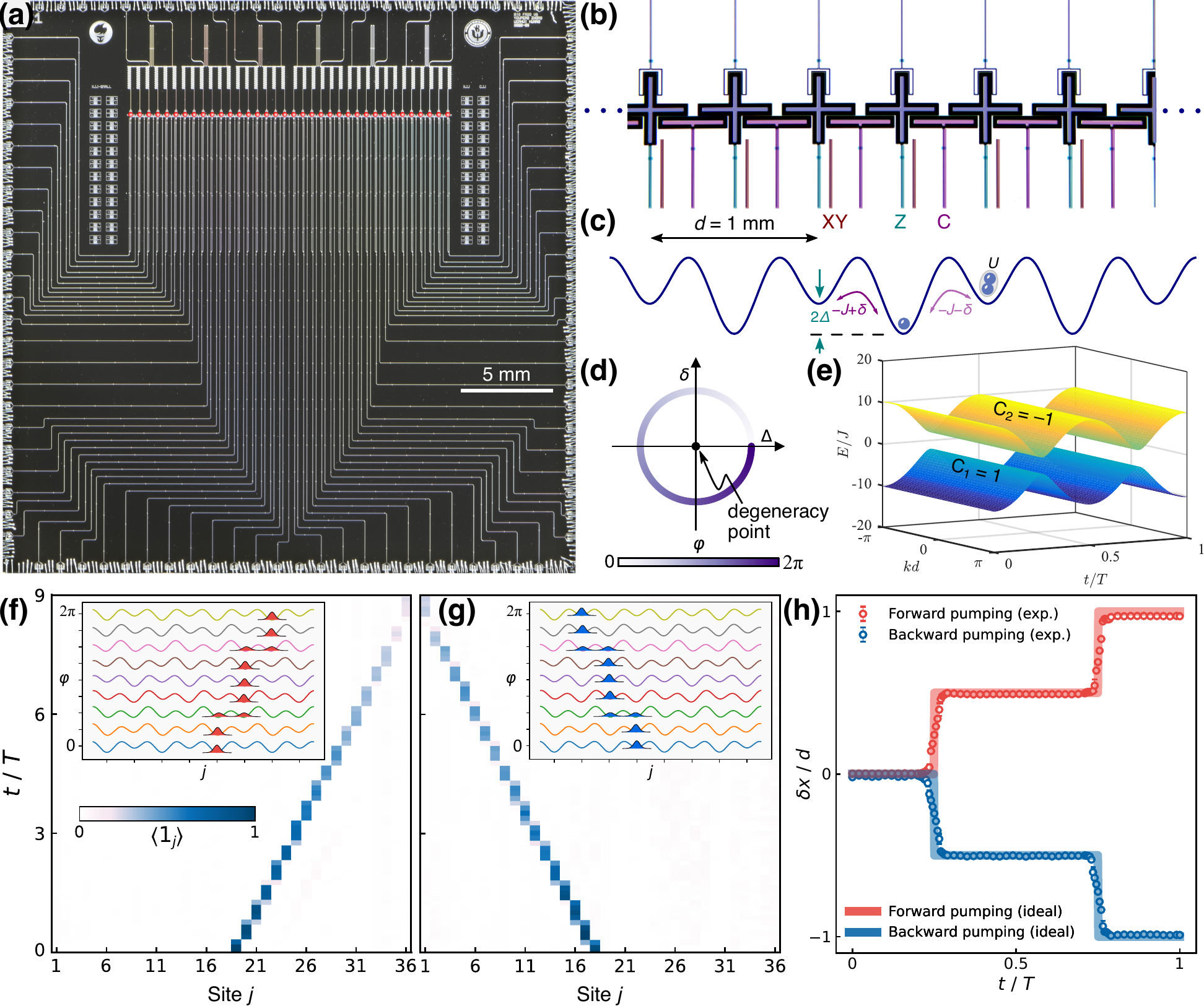}
			\caption{
				\label{fig1}
				{Topological Thouless pumping in an array of superconducting qubits.}
				(a) Photograph of the quantum chip comprising a 1D array of 36 superconducting qubits (marked by red dots).
				(b) Zoomed in micrograph of the chip in (a), showing the X-shaped qubits (light blue) connected with T-shaped tunable couplers (light purple), with their associated control/readout wirings.
				(c) Potential landscape of the dimerized qubit array with a lattice constant of $d=1$~mm, where $2\Delta$ is the double-well potential bias determined by the qubit frequencies (controlled by Z lines), $-J\pm \delta$ is the nearest-neighbour hopping strength determined by the tunable couplers (controlled by C lines), and $U$ is the on-site interaction determined by the anharmonicity of the qubits. Particles can be injected into each site by their XY lines.
				(d) Illustration of the parameter modulation for implementing Thouless pumping, where the Hamiltonian takes an adiabatic cyclic evolution in the $\Delta$-$\delta$ parameter space around the degeneracy point at the origin with a varying phase $\varphi$.
				(e) The band structure in the $k$-$t$ Brillouin zone, showing two energy bands with Chern numbers $C_1=1$ and $C_2=-1$ respectively.
				(f-g) Forward (backward) pumping where the system is initially prepared in a single-particle Wannier state in the lower (upper) band localized at site $j=19$ ($j=18$). Inset shows the variation of the lattice potential and the evolution of the Wannier state during one cycle.
				({h}) Center-of-mass displacement $\delta x$ extracted from (f) and (g), with $\delta x/d=0.971(5)$ and $-0.990(3)$ respectively in a pumping cycle, consistent with the Chern number.
			}
		\end{center}
	\end{figure*}

Following the discovery of quantum Hall effects (QHE)~\cite{Klitzing1980} and topological insulators~\cite{Hasan2010}, topological states of matter have become one of the most active and productive research areas in modern physics.
Thouless pumping~\cite{Thouless1983,Citro2022} is one of the simplest manifestations of topology in periodically driven quantum systems, which shares the same origin of integer QHE and supports quantized transport of noninteracting particles in one dimension.
The topological nature of Thouless pumping makes it robust against modest perturbations such as disorder or interaction~\cite{Niu1984}, and has generated widespread interest for its potential applications, such as current standards~\cite{Niu1990,Pekola2013} and quantum state transfer~\cite{Hu2020}.
While Thouless pumping remains elusive in electron-based condensed matter systems, it has been recently realized in synthetic systems featuring versatility and controllability,
including ultracold atoms~\cite{Wang2013,Nakajima2016,Lohse2016,Nakajima2021,Kao2021,Dreon2022},
photonic waveguides~\cite{Jurgensen2021,Sun2022,Cheng2022a,Jurgensen2023},  acoustic waveguides~\cite{Yang2015,You2022},
and has also been extended to higher dimensions~\cite{Lohse2018,Zilberberg2018} and momentum space~\cite{Ho2012}.
Although the integer QHE, topological insulators and Thouless pumping in noninteracting systems have been well understood, some exotic topological states of matter merely exist due to inter-particle interactions\cite{Rachel2018},
such as the fractional QHE~\cite{Tsui1982,Laughlin1983}, topological Mott~\cite{Raghu2008} and Kondo insulators~\cite{Dzero2010}.
As the basis for generating multi-particle quantum correlations, inter-particle interaction plays a crucial role in collective quantum phenomena, quantum phase transitions, and quantum information processing.
Recently, topological pumping in interacting quantum many-body systems has garnered significant interest~\cite{Tangpanitanon2016,Ke2017,Hayward2018,Kuno2020,Lin2020,Chen2020,Bertok2022,Mostaan2022,Fu2022},
followed by pioneering experiments with nonlinear photonic waveguides~\cite{Jurgensen2021,Jurgensen2023} and ultracold atoms with tunable Hubbard interactions~\cite{Walter2022}.
However, the interplay between topology and multi-particle correlations remains hardly explored experimentally, partly because of the challenge in creating and probing site-resolved multi-particle correlations.

In recent years, superconducting quantum circuits have emerged as a promising platform for exploring interaction effects, such as synthetic many-body interactions~\cite{Zhang2022a}, strongly correlated quantum walks~\cite{Yan2019,Gong2021b}, chiral ground-state currents~\cite{Roushan2017}, antisymmetric spin exchange~\cite{Wang2019}, and multi-particle bound states~\cite{Morvan2022}.
Here we demonstrate interaction-induced topological pumping in such a solid-state quantum system, where a one-dimensional (1D) array of 36 superconducting qubits act as artificial atoms.
With strong interaction inherent in the qubits and site-resolved controllability of the lattice potential and hopping strength, we realize the topological Thouless pumping of single and two bounded particles (microwave photons), in which their center-of-mass positions show quantized transport.
Furthermore, we also observe topologically resonant tunneling and asymmetric edge-state transport of interacting particles when the neighboring-well potential bias matches the interaction energy.
These emergent topological behaviors have no linear or noninteracting counterparts.
The introduction of inter-particle interaction in such fully controlled quantum many-body systems allows for experimental exploration of the interplay between topology and correlations~\cite{Preiss2015,Yan2019}, opening a new avenue for studying exotic topology and transport in interacting quantum many-body systems~\cite{Ozawa2019} and quantum information transfer~\cite{Hu2020} using superconducting quantum circuits.

	\begin{figure*}[t]
		\begin{center}
			\includegraphics[width=0.9\textwidth]{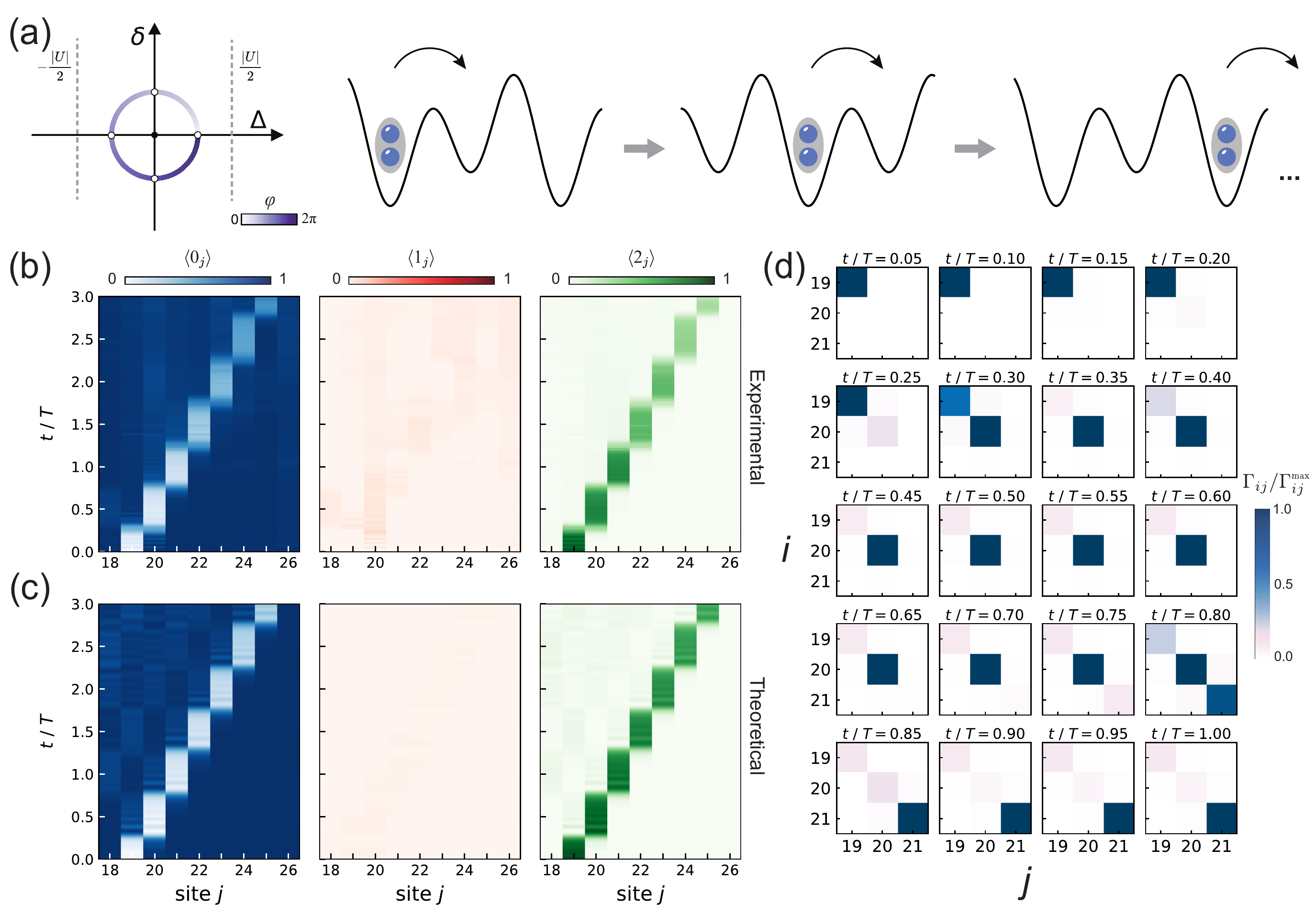}
			\caption{
				\label{fig2}
				{Topological Thouless pumping of two-particle bound states.}
				(a) Schematic showing two interacting particles unidirectionally transported through the barriers as a bound state.
				Experimental data (b) and numerical simulation (c) of the populations of the $|n_j\rangle$ states for $n_j=0$, $1$ and $2$ of each site $j$ during the pumping process, where the $|1_j\rangle$ state is unpopulated.
				(d) The normalized two-site correlations $\Gamma_{ij}/\Gamma_{ij}^{\rm{max}}$ of site $i$ and $j$ during the pumping at different evolution times $t=0.05T$, $0.1T$, $\cdots$, $T$. The probability of finding particles at sites $i$ and $j$ are concentrated along the diagonal of the correlation, manifesting the bound state nature.}
		\end{center}
	\end{figure*}
	
	Our experiments are implemented on a superconducting quantum processor shown in \rfig{fig1}{(a)}, comprising a 1D array of $N=36$ tunably coupled transmon qubits of the Xmon variety~\cite{Barends2013} (see \rfig{fig1}(b)).
	The ability to tune both the lattice potential of each site, as well as the nearest-neighbour hopping strength in a dynamic way offers new opportunity for preparation of topological and otherwise exotic phases of synthetic matter in such systems~\cite{Ma2019,Xu2020,Guo2021,Fedorov2021,Mi2021,Li2022,Xiang2022,Saxberg2022,Mi2022}.
	Here, we configure the system as a dimerized chain with alternating hopping amplitudes $-J\pm \delta$ and staggered potential energies $\pm \Delta$, as shown in \rfig{fig1}(c).
	The physics of this system is captured by the Rice-Mele Hamiltonian ~\cite{Rice1982} with on-site inter-particle interactions,
	\begin{equation}\label{RMH}
		\begin{split}
			H_{RM}/\hbar =\sum_{j=1}^{N-1} \left[[-J + (-1)^j \delta] a_j ^\dag a_{j+1} + \textrm{H.c.}\right] \\ 
			+ \sum_{j=1}^{N} [(-1)^j \Delta n_j + \frac{U}{2} n_j (n_j-1)],
		\end{split}
	\end{equation}

	where $a_j$ is the annihilation operator for site $j$ ranging from 1 to 36, $n_j = a_j^\dag a_j$ is the particle number operator, $-J\pm\delta$ is the nearest-neighbour hopping strength determined by the tunable couplers, $\pm\Delta$ is the staggered on-site energy determined by the qubit frequencies, and $U/2\pi\approx -190$~MHz is the on-site Hubbard interaction strength determined by the anharmonicity of the qubits (see \rfig{fig1}(c)).
	A pair of neighbouring qubits constitutes a unit cell with a lattice constant of $d=1$~mm.
	When $\Delta=0$, this Hamiltonian reduces to the Su-Schrieffer-Heeger (SSH) model~\cite{Su1979} with on-site interactions.
	Leveraging our site-resolved tuning capabilities, we individually excite only particular qubits and implement the Thouless pumping by periodically modulating the qubit frequency and the coupling strength as $\Delta = \Delta_0 \cos (\omega t)$, $\delta = \delta_0 \sin (\omega t)$ respectively, where $t$ is the evolution time, and the pump cycle is given as $T = 2\pi/\omega$.
	The time-evolution can be described in terms of a trajectory $\mathcal{C}$ in the $\Delta$-$\delta$ parameter plane with a varying phase $\varphi=\omega t$, see \rfig{fig1}(d).
	Because the lattice potential is periodic both in space and time, under periodic boundary condition, one can define the Bloch wavefunction $|\psi_{m,k}(t)\rangle=e^{ikx}|u_{m,k}(t)\rangle$ in the $m$-th Bloch band with quasimomentum $k$, and the corresponding topological invariant known as Chern number in a $k$-$t$ Brillouin zone:
	\begin{equation}\label{Chern}
		C_m = \frac{1}{2\pi}\int_0^T dt \int_{-\pi/d}^{\pi/d} dk \Omega_m(k,t),
	\end{equation}
	where $\Omega_m(k,t)=i(\langle \partial_t u_{m,k}| \partial_k u_{m,k}\rangle - \langle \partial_k u_{m,k}| \partial_t u_{m,k}\rangle)$ is the Berry curvature (see Supplementary Information) and $T$ is the pumping period.
	\rfig{fig1}(e) shows a representative instantaneous spectrum $E/J$ of the Hamiltonian in the $k$-$t$ Brillouin zone, with a lower band and an upper band.
	As long as the bandgap never closes, ideally the particles will stay within the same band during the adiabatic pumping process.
	Since the time-reversal symmetry is broken by the phase sweep, the lower and upper band have nontrivial Chern numbers of $C_{1}=1$ and $C_{2}=-1$ respectively.
	The center-of-mass shift of the particles in such topologically nontrivial bands after one pumping cycle is simply given by $\delta x \equiv x(t) - x(0)= C_m d$, where the position operator $x=d/2\sum jn_j/\sum n_j$.

	The instantaneous spectrum of the Hamiltonian becomes gapless at $\delta=\Delta=0$ and as long as the parameter trajectory $(\delta,\Delta)$ encircles this degeneracy point, the pumped charge is quantized.
	In \rfigs{fig1}(f)-\ref{fig1}(g), given parameters $\Delta_0/2\pi=80$~MHz, $\delta_0/2\pi=J/2\pi=8$~MHz, $T=0.4$~$\mu$s, we demonstrate forward (backward) pumping, when the system is initially prepared as a single-particle Wannier state uniformly filling the lower (upper) band.
	The Wannier state is approximately localized at site $j=19$ ($j=18$) by applying a $\pi$ pulse to excite the qubit from the ground state $|0_j\rangle$ to the first excited state $|1_j\rangle$.
	\rfig{fig1}{(h)} shows the center-of-mass displacement $\delta x$ extracted from \rfigs{fig1}(f) and (g), yielding $\delta x/d=0.971(5)$ and $-0.990(3)$ in a pumping cycle respectively, consistent with the Chern numbers of the lower and upper bands.
	We note that the quantization of the pumped particle is only valid as long as the potential is varied adiabatically, as this phenomenon is not generically robust to non-adiabatic effects despite its topological nature~\cite{Privitera2018}.
	The pumping period of $T=0.4$~$\mu$s chosen here is a balance between adiabaticity and qubit coherence, see details in Supplementary Information.
	
	
	\begin{figure*}[t]
		\begin{center}
			\includegraphics[width=0.9\textwidth]{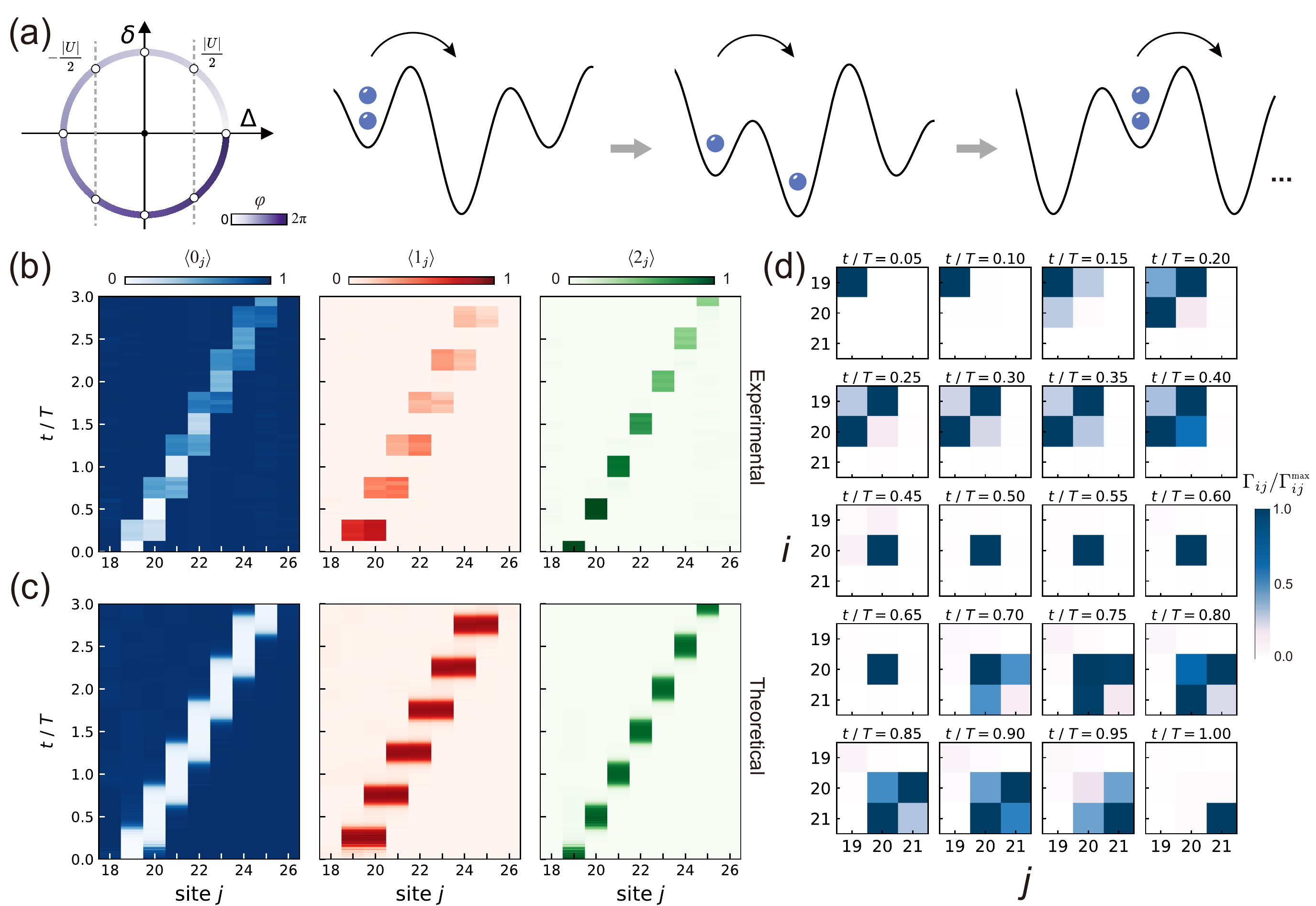}
			\caption{
				\label{fig3}
				{Topologically resonant tunneling of interacting particles.}
				(a) Schematic showing interacting particles transported unidirectionally through the barriers one by the other.
				Experimental data (b) and numerical simulation (c) of the particle populations during the pumping process, where the $|1_j\rangle|1_{j+1}\rangle$ state is populated during the transition between $|2_j\rangle|0_{j+1}\rangle$ and $|0_j\rangle|2_{j+1}\rangle$.
				(d) The normalized correlations $\Gamma_{ij}/\Gamma_{ij}^{\rm{max}}$ during the pumping at different evolution times $t=0.05T$, $0.1T$, $\cdots$, $T$. The off-diagonal elements suggest the two particles occupy two neighboring sites, as a result of the interplay among the inter-particle interaction, the double-well bias and the band topology.
			}
		\end{center}
	\end{figure*}

	When inter-particle interaction is involved, we need to generalize Thouless pumping from single particle to multi-particle cases, where the Bloch wave-function in Eq.~(\ref{Chern}) is replaced with multi-particle Bloch wave-function $|\psi_{m,K}(t)\rangle$ with $K$ being center-of-mass quasimomentum~\cite{Ke2017}.
	The initial two-particle Wannier state approximates to two particles at site $j$ prepared by directly driving the qubit to the second excited state $|2_j\rangle$ via two-photon excitation.
	We focus on the strong interaction regime where $|U| \gg \left|-J\pm \delta\right|$, and vary the double-well bias to explore the transport behavior in different regimes.
	If $2|\Delta_0|< |U|$, the interaction dominates and protects the bound states of two particles in the same site.
	In this case, the two particles in the same site are shifted unidirectionally as a whole~\cite{Ke2017,Lin2020}, see the schematic in \rfig{fig2}(a).
	We perform this two-particle bound-state pumping with $\Delta_0/2\pi=8$~MHz, $\delta_0/2\pi=J/2\pi=12$~MHz, and a pumping period of $T=0.4$~$\mu$s.
	Limited by the qubit decoherence and non-adiabatic effects, we are not able to perform the topological pumping of two-particle bound-states on the full chain.
	We choose a subset of the qubits with index 18--26 to perform this pumping, because these qubits have better readout visibilities and coherence.
	The population of the Fock states $|n_j\rangle$ for $n_j=0$, 1 and 2 of each site $j$ are simultaneously measured and shown in \rfig{fig2}(b), where the $|1_j\rangle$ state of each site is unpopulated.
The corresponding theoretical results are shown in \rfig{fig2}(c), see Supplementary Information for details.
	To manifest the bound-state nature more clearly, we measure the density-density correlations $\Gamma_{ij}=\langle a^\dag_i a^\dag_j a_i a_j \rangle$ of site $i$ and $j$~\cite{Preiss2015,Yan2019} at different evolution times $t$ within a pumping period, as shown in \rfig{fig2}(d). The probability of finding particles at sites $i$ and $j$ are concentrated along the diagonal of the normalized correlation, $\Gamma_{ij}/\Gamma_{ij}^{\rm{max}}$, indicating that the two particles always appear at the same site. This unequivocally demonstrates the bound-state nature of the particles.
	
	If $2|\Delta_0| > |U|$, the double-well bias $2\Delta$ balances the interaction $U$ four times in each pumping cycle, where single-particle resonant tunneling happens and the bound-state is destroyed~\cite{Ke2017}.
	In this case, the two particles in one site are unidirectionally transported through the barriers one by the other, see \rfig{fig3}(a), which is a fundamentally different kind of topological pumping induced by inter-particle interaction.
	In \rfig{fig3}(b), we perform topological pumping with $\Delta_0/2\pi=150$~MHz, an order of magnitude larger than that in \rfig{fig2}(b), while $\delta_0/2\pi=J/2\pi=12$~MHz and $T=0.4$~$\mu$s remain unchanged. The corresponding theoretical results are shown in \rfig{fig3}(c).
	In this case, the $2|\Delta_0|/2\pi = 300$~MHz is clearly larger than $|U|/2\pi=190$~MHz.
	 Similarly, the population of the Fock states $|n_j\rangle$ of each site are simultaneously measured.
	 Different from that in \rfig{fig2}(b), the $|1_j\rangle|1_{j+1}\rangle$ state is populated during the transition between $|2_j\rangle|0_{j+1}\rangle$ and $|0_j\rangle|2_{j+1}\rangle$.
	We measure the density-density correlations $\Gamma_{ij}$ at different evolution times $t$ within a period, see \rfig{fig3}(d), where the strong correlations observed in diagonal elements suggest the two particles appear at the same site.
	Furthermore, strong correlations are also observed in off-diagonal elements at specific times, suggesting the two-particle bound states are destroyed and split into single particles in two sites.
	The topologically resonant tunneling, i.e., the two particles in one site are unidirectionally transported through the barriers one by the other, is a result of the interplay among the inter-particle interaction, the double-well bias, and the band topology, which does not have a noninteracting (or linear) counterpart~\cite{Ke2017}.
	Without the inter-particle interaction, the quantized and unidirectional transport will disappear~\cite{Ke2017}.
	Note in both cases in \rfig{fig2} and \rfig{fig3}, particles are unidirectionally transported through integer cells in one pumping cycle.

	\begin{figure*}[t]
		\begin{center}
			\includegraphics[width=0.9\textwidth]{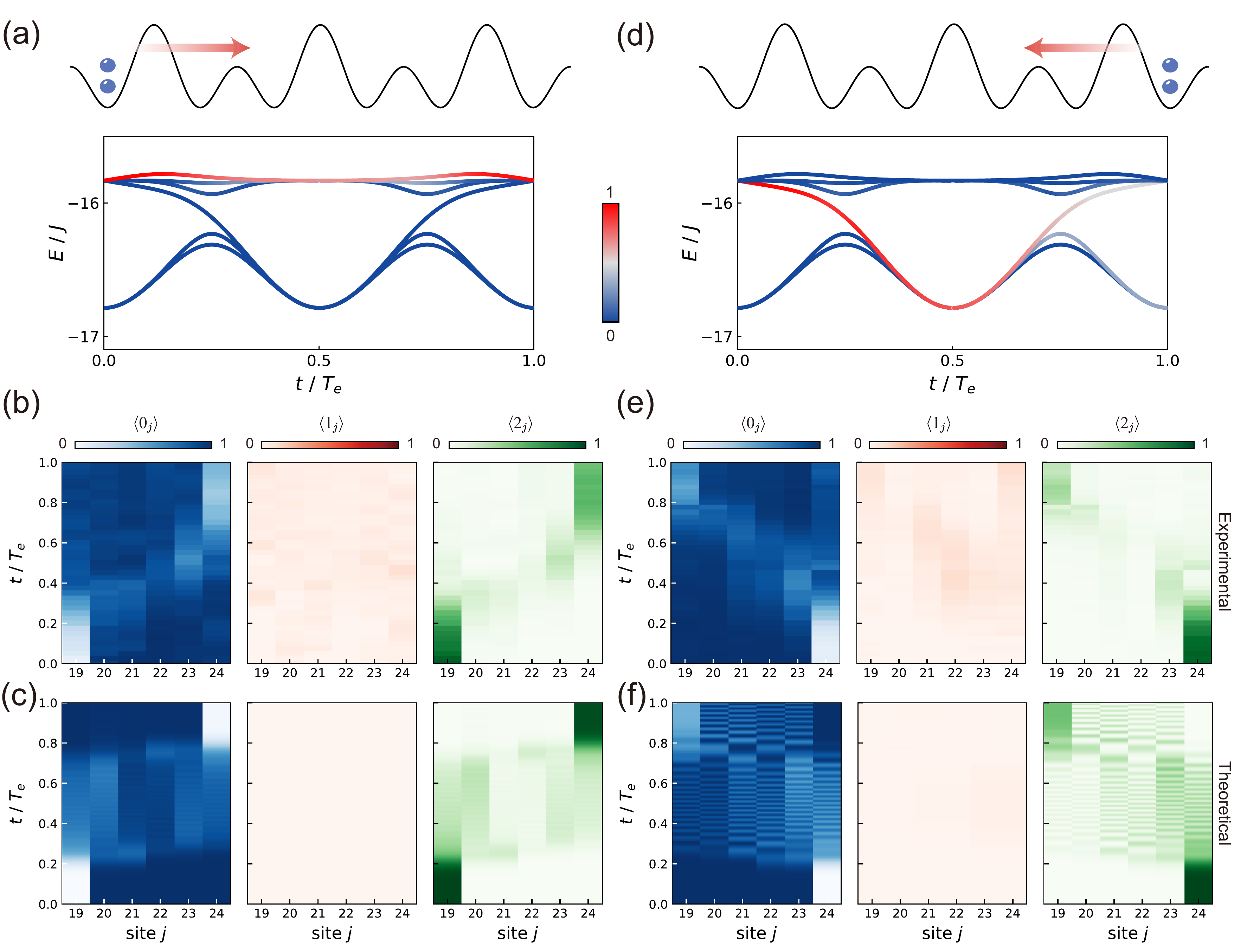}
			\caption{
				\label{fig4}
				{Asymmetric edge-state transport of interacting particles.}
				(a) The evolution of the left edge state in the potential well and the corresponding energy band.
				Experimental data (b) and numerical simulation (c) of the left edge state transport.
				(d) The evolution of the right edge state in the potential well and the energy band.
				Experimental data (e) and numerical simulation (f) of the right edge state transport.
                The colors in (b)-(f) denote the Fock state populations, whereas the colors in the energy band in (a) and (d) denote
                the probability of instantaneous states occupying the eigenstates.
			}
		\end{center}
	\end{figure*}
	
	Finally, we explore the effect of inter-particle interaction to the transport of edge states~\cite{Gorlach2017,Olekhno2020}.
	The topological pumping of edge states in Rice-Mele model (Eq.~\ref{RMH}) can be realized by adiabatically modulating the hopping amplitudes and on-site energy as
		$\Delta=\Delta_0 \sin\left(2\pi t/T_e\right)$, $\delta= -\delta_0 \cos \left(2\pi t/T_e\right)$, where $T_e$ is the pumping period for the edge states.
	In the single particle case, there are two zero-energy instantaneous eigenstates localized at the left (right) edge of the chain under open boundary conditions when $t=0$,
	which become delocalized and immersed into the bulk states during the adiabatic cycle, and finally evolves into the right (left) edge at $t=T_e$.
	The energy spectrum of single-particle Rice-Mele model is symmetric about zero, which provides reversible channels of topological pumping from left to right edge or vice verse, see Supplementary Information for details.
	When inter-particle interaction is involved, the symmetry of energy spectrum is broken as shown in \rfigs{fig4}(a),(d),
	which gives a flat (steep) energy band for the topological pumping starting from the left (right) edge, and thus the non-adiabatic effects of right-to-left pumping are stronger than left-to-right pumping for a given period $T_e$.
	We perform this interaction-induced topological pumping from left (right) edge in a subset of qubits with index 19-24 isolated from other qubits, with
		$\Delta_0/2\pi = 0.5$~MHz, $\delta_0/2\pi = J/2\pi = 12$~MHz
		and $T_e=4~\mu$s, an order of magnitude larger than $T$.
	In \rfigs{fig4}(b),(e), the population of the Fock states $\vert n_j\rangle$ for $n_j=0$, 1 and 2 of each site $j$ are simultaneously measured,
		where the state $\vert 2_{j=19} \rangle$ ($\vert 2_{j=24} \rangle$) is prepared at $t=0$,
		and transported into the right (left) edge of site $j=24$ ($j=19$) at $t=T_e$.
The corresponding theoretical results are shown in \rfigs{fig4}(c),(f).
	
	In conclusion, we have demonstrated the interaction-induced topological pumping in a solid-state quantum system comprising an array of 36 superconducting qubits.
	Our results clearly show that inter-particle interactions can induce emergent quantized transport and topological behaviour with no counterpart in the noninteracting (or linear) case, allowing for the exploration of interplay between band topology and multi-particle correlations~\cite{Preiss2015,Yan2019}.
	These experiments open up a new avenue for studying exotic topological phenomena and information transfer in synthetic quantum many-body systems~\cite{Ozawa2019}.
	Our demonstration could also be extended to two-dimension using flip-chip packaged superconducting qubits~\cite{Mi2021}, enabling the realization of an analogy of the 4D integer quantum Hall effect~\cite{Lohse2018,Zilberberg2018}.
	
	\clearpage
	
	\bibliography{pump}
	
	\section*{Acknowledgments}
	We thank Xiongjun Liu, Yucheng Wang, Dawei Lu and Benchuan Lin for helpful discussions.
	{\bf Funding:} {This work was supported by the Key-Area Research and Development Program of Guangdong Province (2018B030326001), the National Natural Science Foundation of China (U1801661, 12174178, 11905098, 12204228, 12025509 and 12275365), the Guangdong Innovative and Entrepreneurial Research Team Program (2016ZT06D348), the Guangdong Provincial Key Laboratory (2019B121203002), the Science, Technology and Innovation Commission of Shenzhen Municipality (KYTDPT20181011104202253, KQTD20210811090049034, K21547502), the Innovation Program for Quantum Science and Technology (2021ZD0301703), the Shenzhen-Hong Kong Cooperation Zone for Technology and Innovation (HZQB-KCZYB-2020050), the NSF of Beijing (Z190012), and the National Key Research and Development Program of China (2022YFA1404104).}
	{\bf Author contributions:} {
	D.Y. supervised the project.
	C.L., Y.K., Z.T. and X.G. conceived the idea.
	The devices were designed by W.H. and fabricated by L.Z. under the supervision of S.L.
	Z.T., W.H. and J.N. performed the measurements and analyzed the data under the supervision of Y.C. and Y.Z.
	Jiawei Z. and Y.Z. built the custom electronics for qubit control and readout.
	All authors contributed to discussions and production of the manuscript.
	}
	{\bf Competing interests:} The authors declare no competing interests.
	{\bf Data and materials availability:} The data that support the plots within this paper and other findings of this study are available from the corresponding authors upon reasonable request.
	
	


	\clearpage

\end{document}


	\title{Supplementary Information for ``Interaction-induced topological Thouless pumping''}
	
	
	\setcounter{equation}{0}
	\setcounter{figure}{0}
	\setcounter{table}{0}

	\renewcommand{\theequation}{S\arabic{equation}}
	\renewcommand{\thefigure}{S\arabic{figure}}
	\renewcommand{\thetable}{S\arabic{table}}

	\newpage
	
	\section{Experimental setup}
	\begin{figure}[H]
		\centering
		\includegraphics[width=0.95\textwidth]{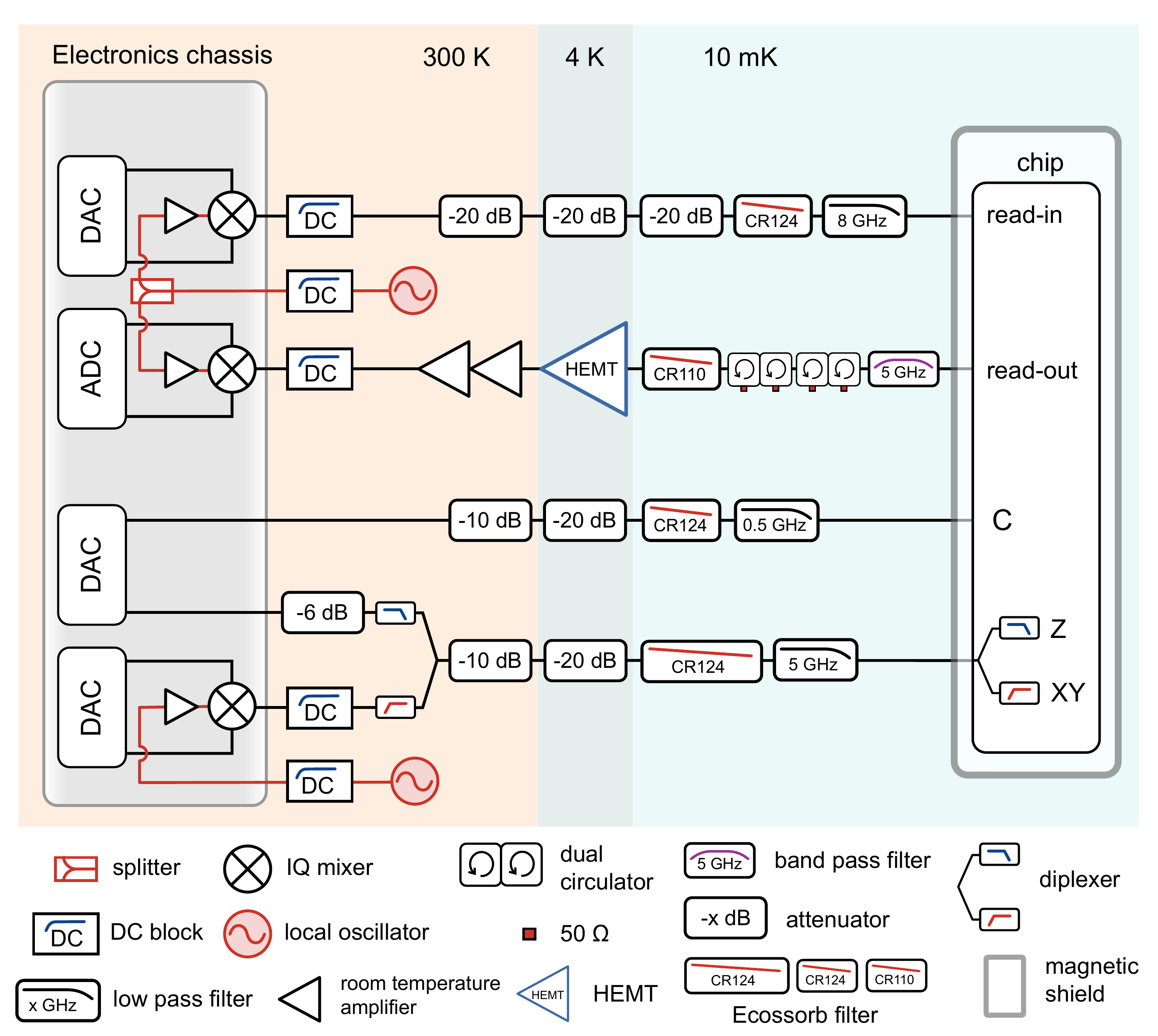}
		\caption{Room temperature and cryogenic wiring.
		}\label{wiring}
	\end{figure}
	
	\rfig{wiring} shows the room temperature and cryogenic wiring layout. We use custom made digital-to-analog converter (DAC) and analog-to-digital converter (ADC) circuit boards for qubit control and measurement, respectively. The control boards have dual-channel 14-bit vertical resolution DAC integrated circuits operating at 1~Gs/s driven by a field-programmable gate array (FPGA) chip. Each DAC analog output is filtered by a custom Gaussian low-pass filter with 250 MHz bandwidth to filter the clock feedthrough, where the Gaussian spectrum of the filter ensures smooth impulse response in time domain.
	The DAC boards can generate nanosecond-length pulses for fast qubit $Z$ or coupler $C$ control, or the two channels can modulate the in-phase and quadrature components of an IQ mixer for frequency up-conversion, providing several GHz frequency signals for qubit XY control and dispersive readout.
	At room temperature, the XY and Z control signals of each qubit are combined together with a diplexer and delivered to the device using one coaxial cable channel.
	The combined control signals are then split using an on-chip diplexer~\cite{Niu2023}, in which the XY signal is capacitively coupled to the qubit via an open-ended quarter-wavelength impedance matching line, while the Z signal is inductively coupled to the qubit through another short-ended quarter-wavelength impedance matching line.
	
	The ADC boards have dual-channel 8-bit vertical resolution ADC integrated circuits operating at 1~Gs/s driven by a FPGA chip.   
	The qubit dispersive readout output is first amplified by the cryogenic high electron mobility transistor (HEMT) manufactured by Low Noise Factory at the 4~K stage and two room-temperature low noise amplifiers, 
	then downconverted with an IQ mixer, and finally digitized and demodulated by the  ADC boards.
	Two cryogenic circulators are inserted between the device and the cryogenic HEMT to isolate reflections as well as thermal noise emitted from the input of the cryogenic HEMT to the device.
	The Eccorsorb CR110 and 5 GHz bandpass filters with low insertion loss at output signal frequency are used to block noise outside the bandwidth of circulators.
	Each control line is heavily attenuated and filtered at each temperature stage in the dilution refrigerator to minimize the impact on the qubit coherence while retaining controllability.

\begin{figure}[H]
		\centering
		\includegraphics[width=0.98\textwidth]{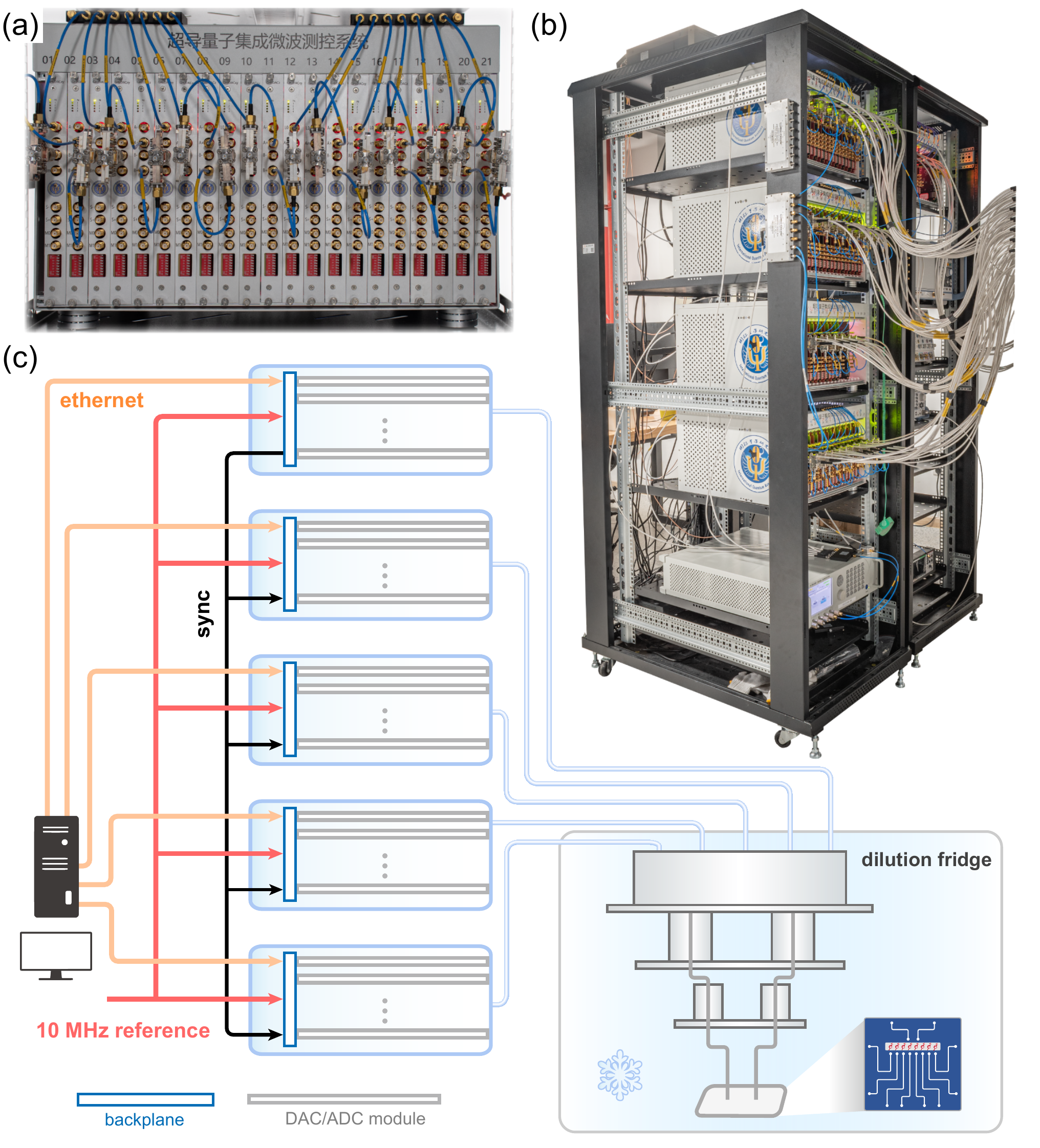}\\
		\caption{Microwave electronics for qubit control and readout.
			(a) Photograph of a 5U chassis of custom made DAC/ADC modules.
			(b) Photograph of the microwave electronics setup housed in two 42U racks, consisting of 5 chassis of custom made DAC/ADC modules, together with microwave signal generators, splitters and diplexers etc.
			(c) Schematic of the electronics connection, showing the 10 MHz reference clock, the synchronization and the ethernet network topology.
		}\label{electronics}
	\end{figure}

	The custom DAC/ADC boards receive power, communication, and clock synchronization through a custom backplane housed in a 5U chassis, which can host 21 DAC/ADC boards, see \rfig{electronics}(a). 
	In our experiments, we used up to 84 boards, including 42 DAC boards for XY control, 36 DAC boards for Z control, and 6 ADC boards for data acquisition, all hosted in five chassis as illustrated in \rfig{electronics}(b). 
	The chassis are synchronized with a 10 MHz reference clock, and the output signals of all boards are synchronized  by trigger signals generated by a master board and distributed to other boards in the same chassis through the backplane, and distributed to other four chassis through coaxial cable connections.
	Communication between the host computer and the chassis is through a local area ethernet network. The electronics connection topology is depicted in \rfig{electronics}(c).

	\clearpage
	
	\section{Device information}
	\subsection{Device fabrication}
	The superconducting quantum processor used in this experiment is fabricated with the following steps:
	\begin{enumerate}
		\setlength{\itemsep}{0pt}
		\item A 100 nm aluminum film is deposited on a 4 inch c-plane sapphire wafer in Plassys MEB 550SL3.
		\item Photo-lithography using laser direct writing, followed by inductively coupled plasma (ICP) dry etching to define the capacitor pads, readout resonators, control and readout circuits etc.
		\item Photo-lithography using laser direct writing, followed by electron beam evaporation of 200 nm SiO$_2$ and lift-off to fabricate the scaffold layer for air-bridge crossovers.
		\item Electron-beam lithography, followed by double-angle evaporations of aluminum in Plassys MEB 550SL3 and lift-off to fabricate the Al/Al$_2$O$_3$/Al Josephson junctions.
		\item Photo-lithography using laser direct writing, followed by electron beam evaporation of 300 nm aluminum and lift-off to fabricate the conducting layer for air-bridge crossovers and bandage for the Josephson junctions~\cite{Dunsworth2017,Dunsworth2018}, with \emph{in situ} ion milling to create galvanic contact with the bottom aluminum layer fabricated in step 1.
		\item Dice the wafer into individual dies.
		\item Release the SiO$_2$ scaffold layer using vapor HF etch.
	\end{enumerate}
	
	\clearpage
	
	\subsection{Device performance}
	
\begin{figure}[!htbp]
		\centering
		\includegraphics[width=0.8\textwidth]{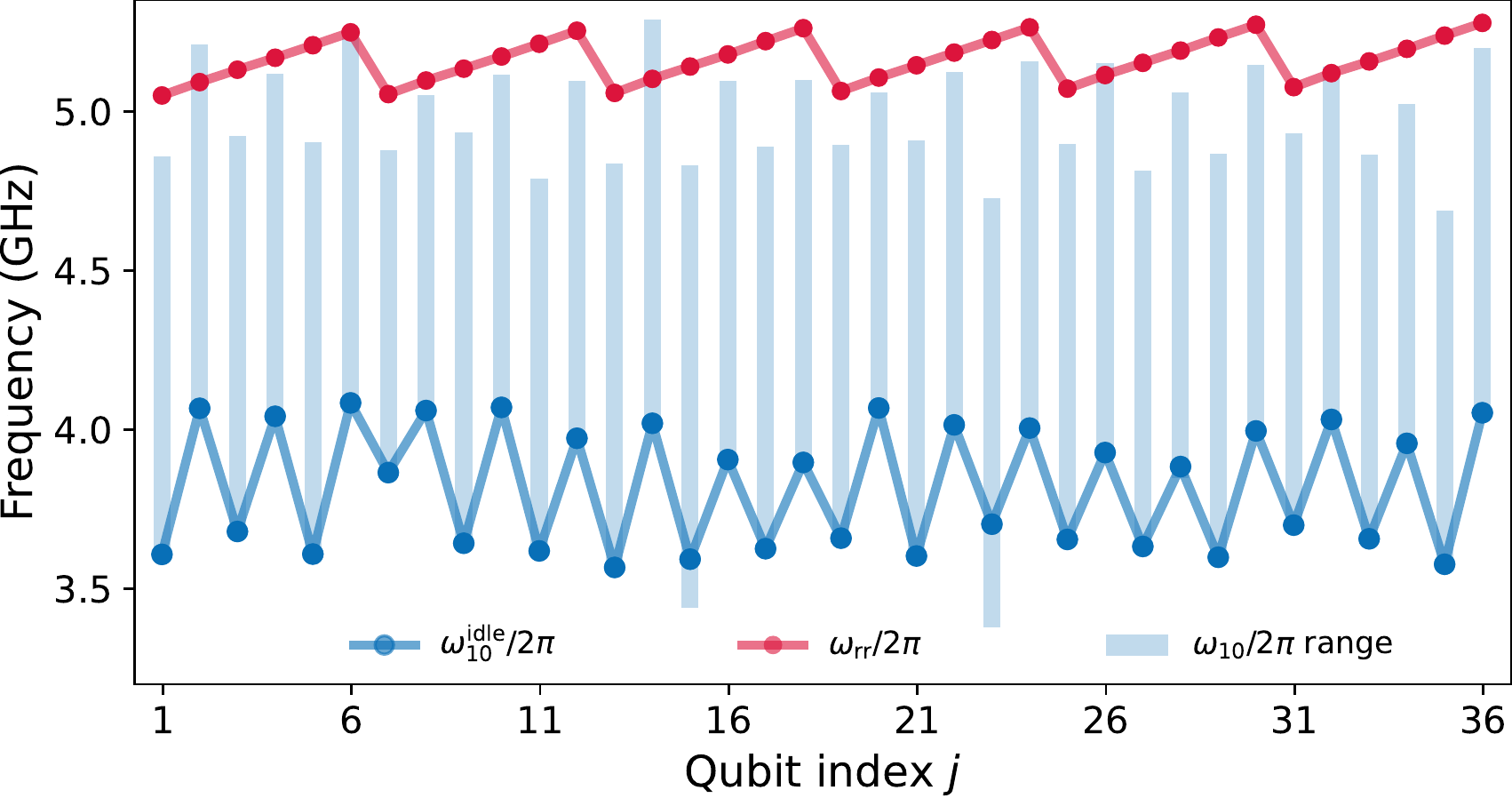}
		\caption{\label{Qpara_qr_freq}
			Qubit frequencies $\omega_{10}/2\pi$ and readout resonator frequencies $\omega_{rr}/2\pi$.
		}
	\end{figure}

	In this experiment, we use asymmetric Josephson junctions with $\alpha=E_{J1}/E_{J2}=3.3~ (4.3)$ on the odd (even) indexed qubits, 
	where $E_{J1}$ and $E_{J2}$ are the Josephson energies of the two qubit junctions. 
	With this configuration, the odd (even) indexed qubits have a maximum frequency sweet spot near $4.9$ GHz ($5.1$ GHz)
	and a minimum frequency sweet spot near $3.6$ GHz ($4.1$ GHz), see \rfig{Qpara_qr_freq}.
	The staggered frequency arrangement of the qubits in the chain enables appropriate idling frequencies for optimal performance.
	The upper sweet point of qubits is close to the readout resonators frequency range of 5--5.3 GHz, allowing for fast reset proposals~\cite{Zhou2021}.
	The 36 readout resonators are divided into six groups, each group is coupled to a Purcell filter~\cite{Jeffrey2014}. The Purcell filter functions as a bandpass filter, impeding microwave
	propagation at the qubit frequency, and suppressing the Purcell decay rate.
	
\begin{figure}[!htbp]
		\centering
		\includegraphics[width=0.95\textwidth]{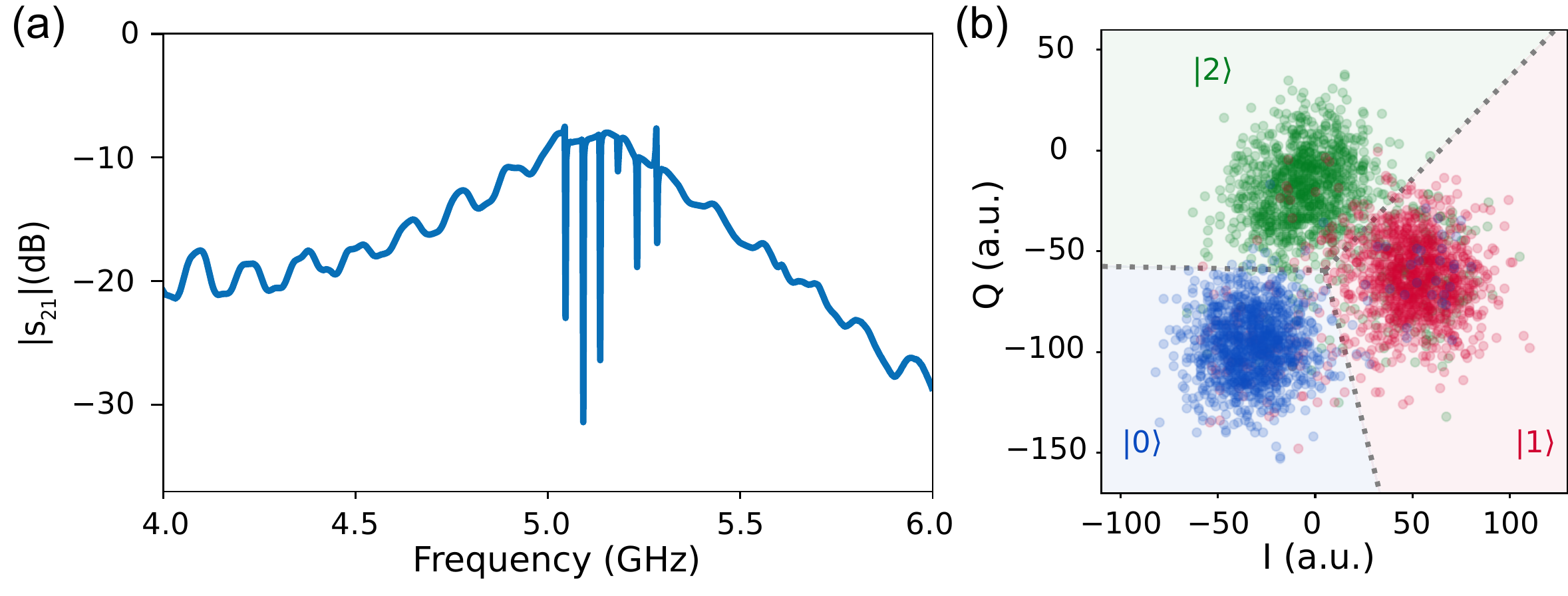}
		\caption{\label{supply_iqraw012_purcell} Single-shot qubit dispersive readout with Purcell filter. 
			(a)
			Measured transmission spectrum of the Purcell filter.
			(b)
			Single-shot dispersive readout for states $\vert 0\rangle, \vert 1\rangle, \vert 2\rangle$ in the quadrature (IQ) space. 
		}
	\end{figure}

	The readout of the qubit state involves probing the resonator's state-dependent frequency shift (dispersive shift), which is coupled but largely detuned from the qubit. However, due to the resonator's spontaneous decay to the readout feed line, the qubit experiences additional damping through the Purcell effect. To mitigate this effect, the above-mentioned Purcell filter can be used as a bandpass filter to impede microwave propagation at the qubit frequency and suppress the Purcell decay rate. The center frequency of the Purcell filter used in our experiment is 5.15 GHz, with weak coupling to the input port and strong coupling to the output port, as seen in its transmission spectrum in \rfig{supply_iqraw012_purcell}(a). By using this Purcell filter, high-quality, single-shot qubit dispersive readout for states $\vert 0\rangle, \vert 1\rangle$ and $\vert 2\rangle$ can be achieved, as demonstrated in \rfig{supply_iqraw012_purcell}(b).
	
\begin{figure}[!htbp]
		\centering
		\includegraphics[width=0.8\textwidth]{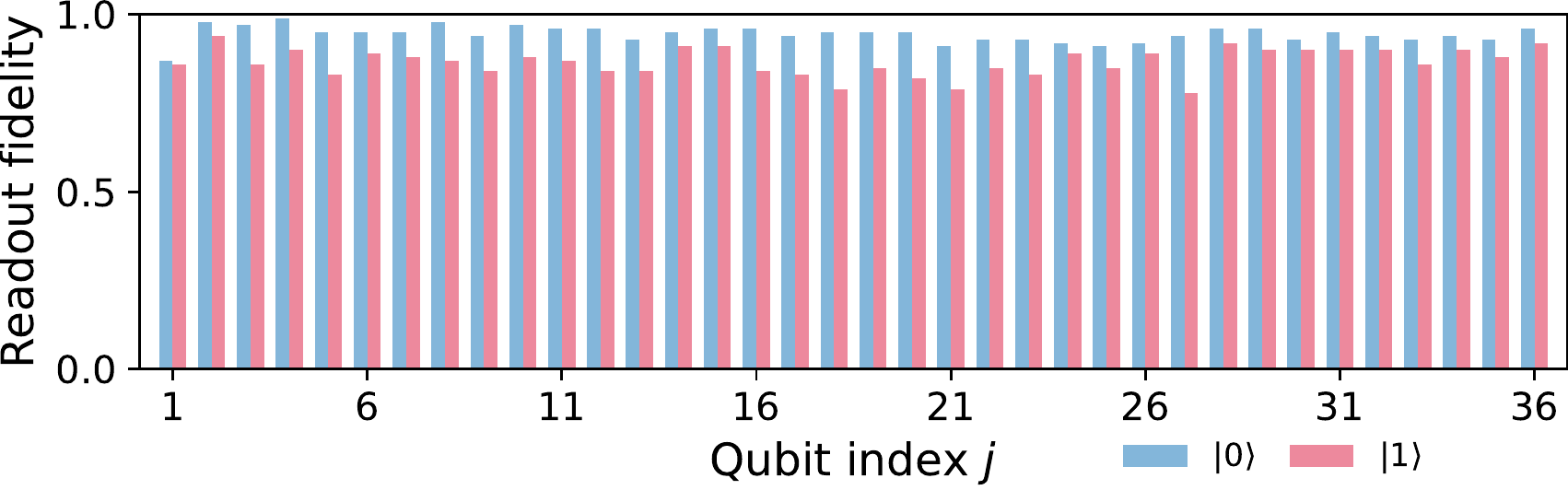}
		\caption{\label{Qpara_readfid} Qubit readout fidelities.
		}
	\end{figure}

	Figure \ref{Qpara_readfid} displays the qubit readout fidelities, with an average state fidelity of 0.95 for the $|0\rangle$ state and 0.87 for the $|1\rangle$ state, respectively.
	
\begin{figure}[!htbp]
		\centering
		\includegraphics[width=0.8\textwidth]{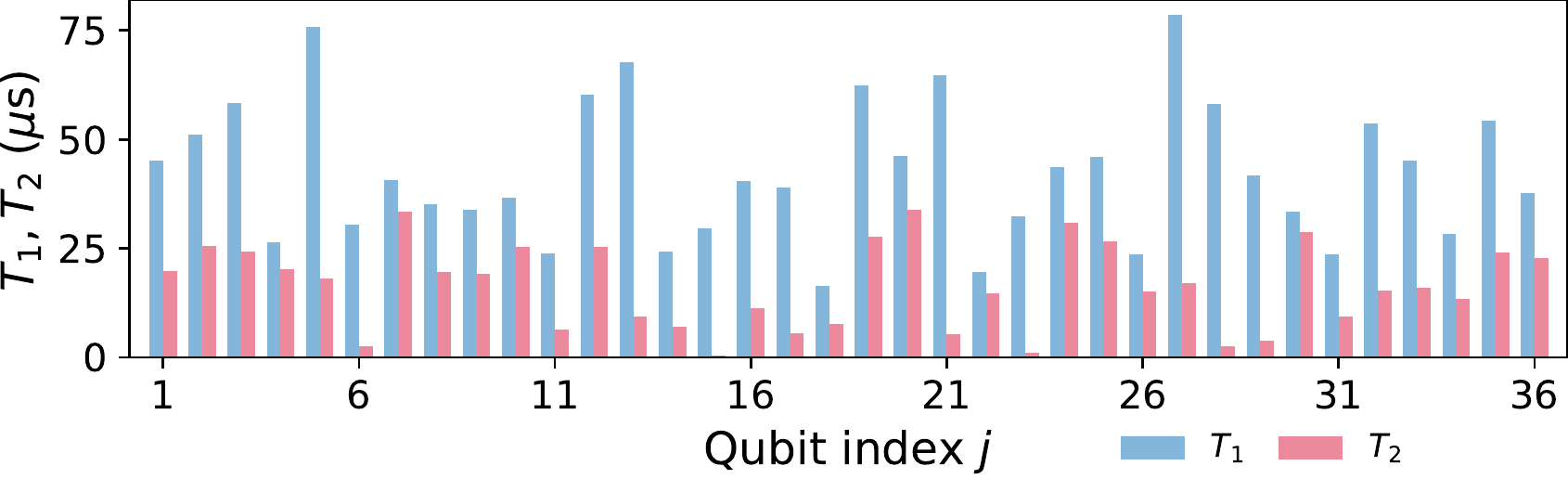}
		\caption{\label{qubit_T1T2} Qubit energy relaxation time $T_1$ and dephasing time $T_2$.
		}
	\end{figure}

	Figure \ref{qubit_T1T2} illustrates the energy relaxation ($T_1$) and dephasing time ($T_2$) of each qubit at its idling frequency. 
	
	\subsection{Device control calibration}
	
	\begin{figure}[htbp]
		\centering
		\includegraphics[width=0.8\textwidth]{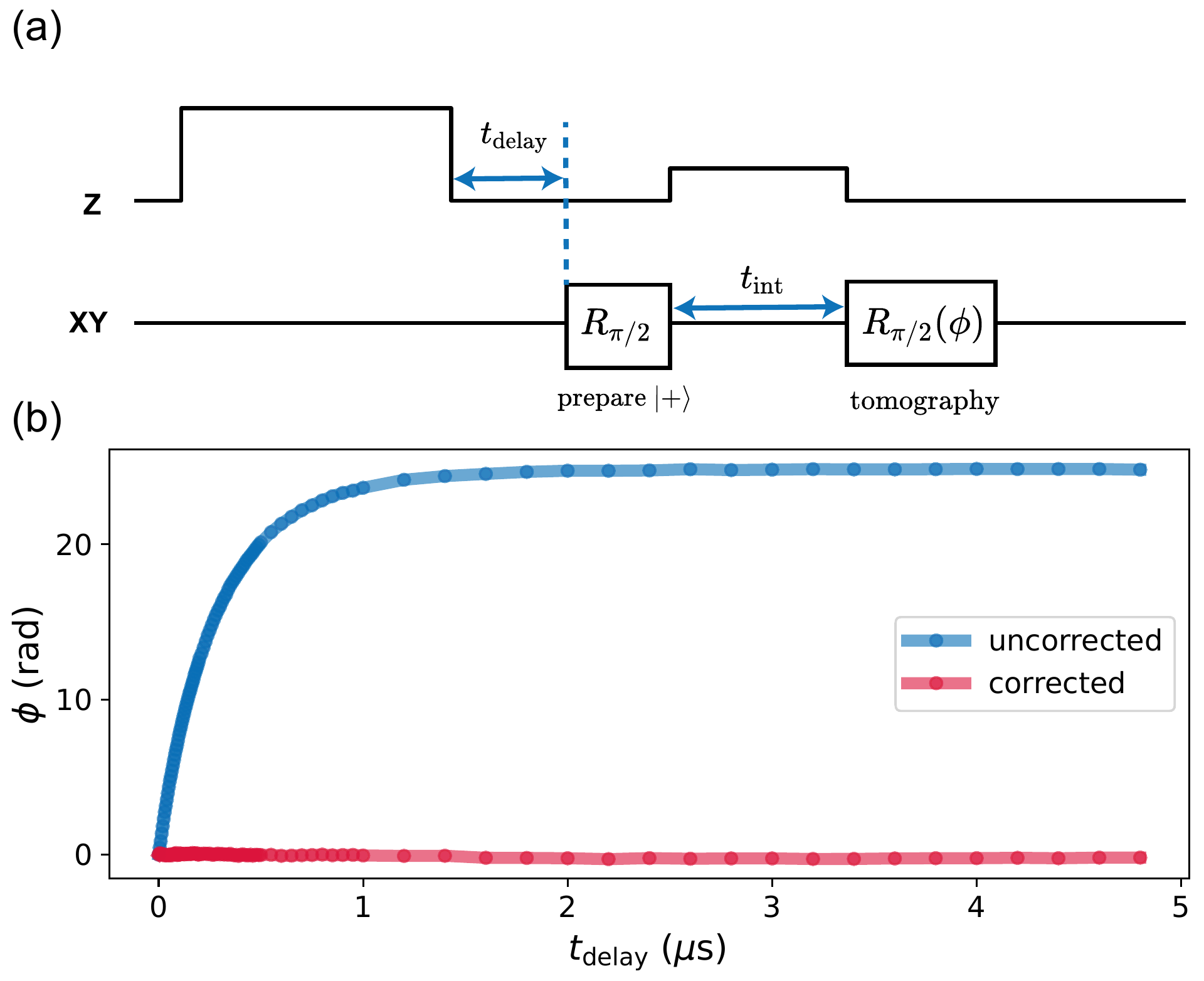}
		\caption{\label{supply_Zdistort} Correction of Z pulse distortion.
			(a)
			The pulse sequence used to measure Z pulse distortion. 
			The first $R_{\pi/2}$ pulse is applied after the Z pulse with a delay time $t_{\mathrm{delay}}$ to prepare 
			$\vert + \rangle = (\vert 0 \rangle + \vert 1 \rangle )/\sqrt{2}$, 
			then the second $R_{\pi/2}(\phi)$ pulse for qubit tomography is applied to obtain the accumulated phase 
			during the integration time $t_{\mathrm{int}}=100~\mathrm{ns}$, 
			where qubit frequency is placed on a phase sensitive point during $t_{\mathrm{int}}$.
			(b)
			The accumulated phase as a function of $t_{\mathrm{delay}}$, where the blue and red curve are for the phases before
			and after the correction of pulse distortion, respectively.
		}
	\end{figure}
	We consider the distorted Z pulse $X_{\mathrm{qubit}}$ in the frequency domain that arrives at the qubits as
	\begin{equation}\label{Zpulse_freqSpace}
		X_{\mathrm{qubit}}(\omega) = H(\omega) X_{\mathrm{in}}(\omega), 
	\end{equation}
	where $X_{\mathrm{in}}(\omega)$ is the input signal, and $H(\omega)$ is the transfer function. 
	To correct this distortion, we perform a predistortion on our desired pulse $X(\omega)$ with the inverse of $H(\omega)$
	\begin{equation}
		X_{\mathrm{corr}}(\omega) = H^{-1}(\omega) X(\omega), 
	\end{equation}
	where
	\begin{equation}
		H(\omega) = 1+ \sum_{j=1}^{k} \frac{ i a_j \omega }{i\omega + b_j}
	\end{equation}
	describes multiple exponentially decaying overshoots ($a_j>0$) or undershoots ($a_j<0$) with decay rate $b_j$. 
	One can identify such distortion by preparing a superposition state 
	$\vert + \rangle = (\vert 0 \rangle + \vert 1 \rangle )/\sqrt{2}$ 
	after applying the Z pulse, and subsequently monitoring the phase  of the second $R_{\pi/2}(\phi)$ pulse.
	\rfig{supply_Zdistort} gives the qubit phase $\phi$ as a function of the delay time $t_{\mathrm{delay}}$ before and after the correction of Z pulse.
	
\begin{figure}[!htbp]
		\centering
		\includegraphics[width=0.8\textwidth]{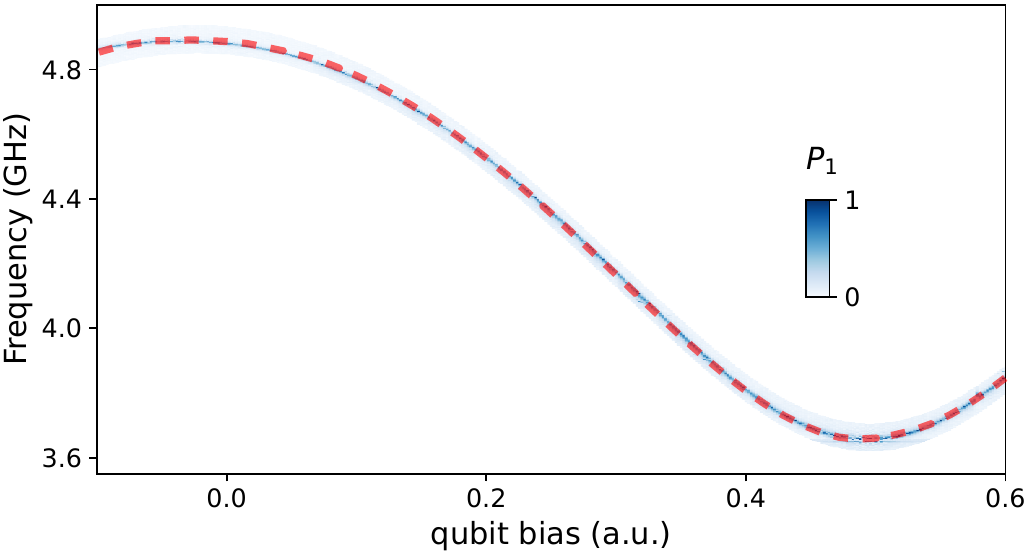}
		\caption{\label{supply_freq_zpa2d} Characterization of the qubit frequency by spectroscopy.
		}
	\end{figure}

	The qubits frequencies can be individually adjusted by Z control lines, 
	where a typical relationship between the qubit frequency and the qubit bias is given in \rfig{supply_freq_zpa2d}.
	
\begin{figure}[!htbp]
		\centering
		\includegraphics[width=0.8\textwidth]{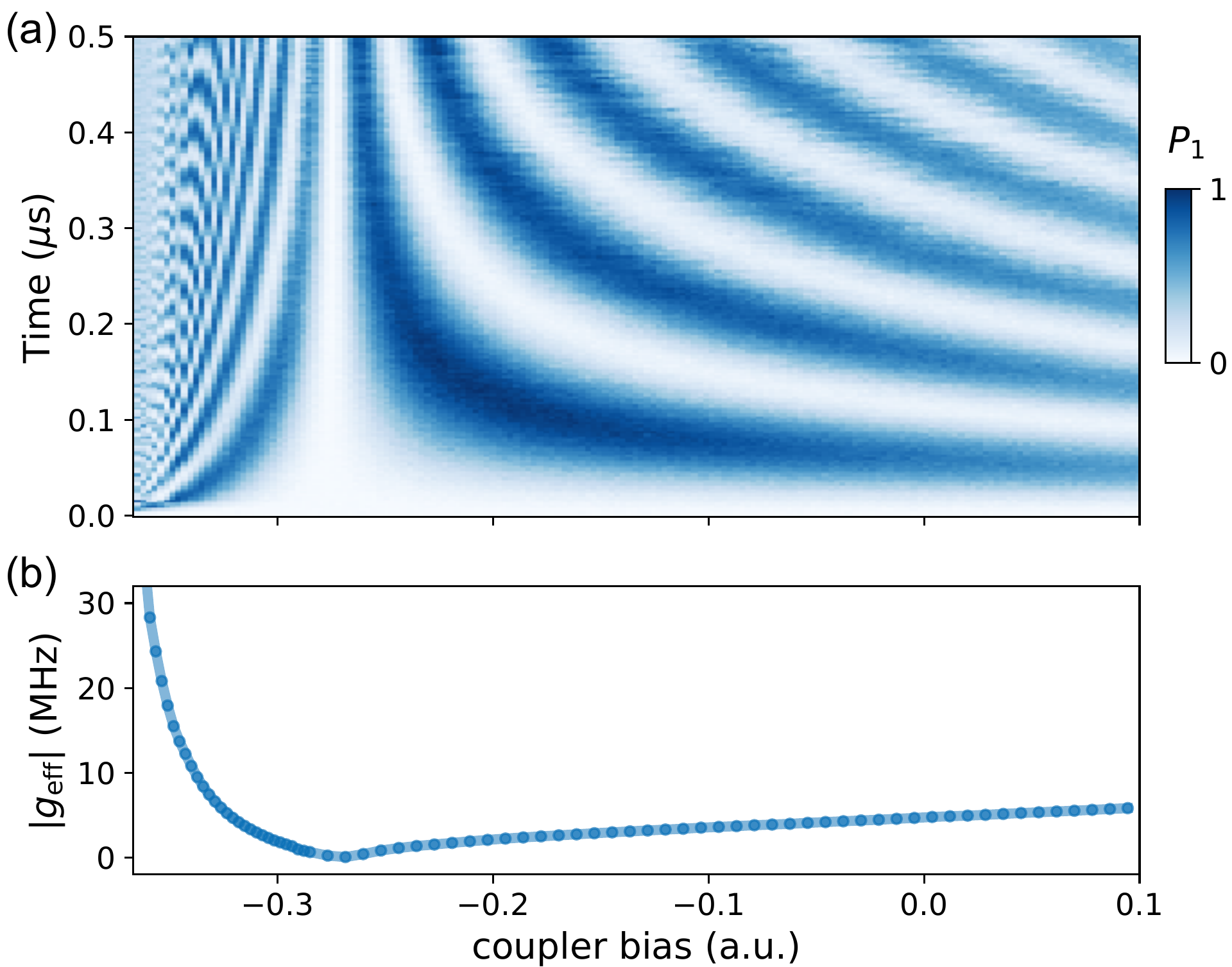}
		\caption{\label{supply_iswap_cpa2d}
			Characterization of the T-shaped transmon coupler $C_{1,2}$.
			(a)
			Vacuum Rabi oscillation between $Q_1$ and $Q_2$ at different coupler bias,
			in which the colors denote the population $P_{1}$ of $Q_2$.
			(b)
			The coupling strength extracted from the data of (a).
		}
	\end{figure}

	In our device, the X-shaped qubits are coupled to each other with a transmon style, T-shaped coupler $C_{j,j+1}$ \cite{Xu2020a}, 
	which provides a tunable coupling strength from $3$~MHz to $-25$~MHz continuously;
	see \rfig{supply_iswap_cpa2d}.

\begin{figure}[!htbp]
		\centering
		\includegraphics[width=0.8\textwidth]{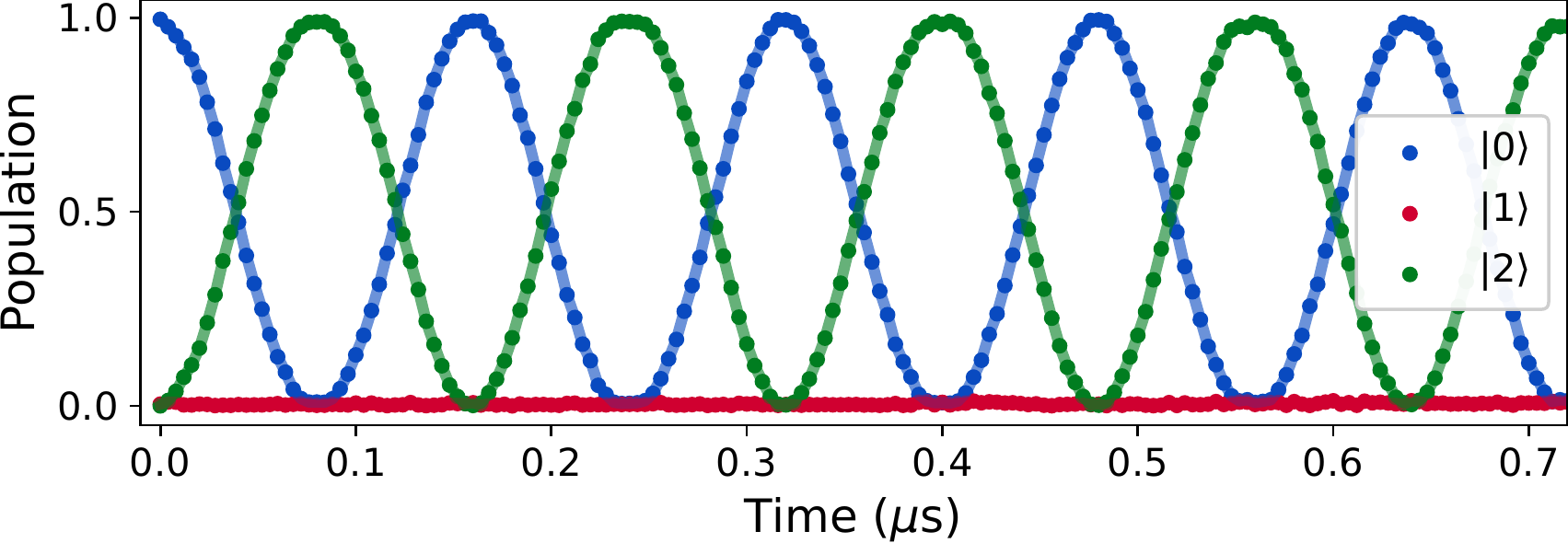}
		\caption{\label{supply_rabi02} {Population of states $\vert 0 \rangle$,  $\vert 1 \rangle$ and $\vert 2 \rangle$  under the microwave driving at frequency $\omega_{02}/2$.}
		}
	\end{figure}

	Figure \ref{supply_rabi02} gives the two-photon Rabi oscillations of $\vert 0 \rangle$ and $\vert 2 \rangle$, 
	where a microwave pulse with frequency $\omega_{02}/2$ is applied through XY control line to perform two-photon excitation\cite{Linskens1996} from $\vert 0 \rangle$ to $\vert 2 \rangle$. 

	\clearpage
	
	\section{Theoretical model and numerical simulation}
	\subsection{Adiabatic shifting of a particle on a dimer}
	To understand the charge pumping protocol from the bottom up, we consider the simplest case with a single dimer only, i.e., two qubits Q1 and Q2 tunably coupled to each other. The Hamiltonian reads
	\begin{equation}\label{Hdimer}
		H_2(t)/\hbar = \Delta(t) ( a^\dag_1 a_1 -   a^\dag_2 a_2) + J(t) (a^\dag_1 a_2 + a_1 a^\dag_2)
	\end{equation}
	In the single excitation manifold, this Hamiltonian can be reduced to a spin Hamiltonian $\widetilde{H_2}/\hbar = \Delta(t)\sigma_z + J(t) \sigma_x$, where $\sigma_z = |10\rangle \langle10|-|01\rangle \langle 01|$, and $\sigma_x = |10\rangle \langle01|+|01\rangle \langle 10|$.
	We initialize the system with a particle at site 1 and then vary the qubit frequencies and hopping strength as $\Delta = \Delta_0\cos(\omega t)$, $J = J_0\sin(\omega t)$.
	This is equivalent to a spin under a slowly rotating magnetic field.
	According to the adiabatic theorem, if $H_2$ is varied slowly enough, the particle will be shifted from $|10\rangle$ to $|01\rangle$ until $\omega t = \pi$.
	
	\subsection{Noninteracting Rice-Mele model}
	The second-quantization Hamiltonian of the system reads
	\begin{equation}\label{RMsecond}
		{H} = \sum_{j,k}{c}_j^\dag H_{j,k} {c}_k
	\end{equation}
	where ${c}_j$ is the creation operator of a particle at site $j$ ($j=1$, 2, $\cdots$, $N$) and ${H}$ is the $N\times N$ matrix of hopping amplitudes in the Wannier basis. For noninteracting Rice-Mele model in the tight binding limit (Equation 1 in the main text), the matrix $H$ takes the form
	\begin{equation}
		H/\hbar=
		\begin{bmatrix}
			\Delta & -J-\delta & 0 & 0 &          \cdots & 0 & 0 & 0 & 0 \\
			-J-\delta & -\Delta & -J+\delta & 0 & \cdots & 0 & 0 & 0 & 0 \\
			0         & -J+\delta & \Delta & -J-\delta &       \cdots & 0 & 0 & 0 & 0 \\
			0         & 0 & -J-\delta & -\Delta &       \cdots & 0 & 0 & 0 & 0 \\
			\cdots    & \cdots & \cdots & \cdots & \cdots & \cdots & \cdots & \cdots & \cdots \\
			0         & 0 & 0 & 0 & \cdots & -J-\delta & -\Delta & -J+\delta & 0 \\
			0         & 0 & 0 & 0 & \cdots & 0 & -J+\delta & \Delta & J-\delta \\
			0         & 0 & 0 & 0 & \cdots & 0 & 0 & -J-\delta & -\Delta \\
		\end{bmatrix}
	\end{equation}
	In the thermodynamic limit ($N\rightarrow \infty$), the Rice-Mele Hamiltonian shows two energy bands separated by a gap with width $2\hbar\sqrt{\delta^2+\Delta^2}$, which vanishes at the degeneracy point $\Delta=\delta=0$.
	
	\subsection{Interacting Rice-Mele model}
	We consider the topological pumping of $N$ interacting particles in a periodically modulated lattice with spatial period $d$.
	The inter-particle interaction breaks the translation symmetry of individual particles.
	However, all the particles shifted as a whole to a unit cell to preserve the energy of the system. 
	Therefore, the system has co-translation symmetry, and the center-of-mass momentum $K$ is a good quantum number~\cite{Ke2017}.
	With the co-translation symmetry, we can apply the multiparticle Bloch theorem to obtain the center-of-mass energy band $E_{m,K}$ and the associated multiparticle Bloch states $|\Psi_{m,K}\rangle$, where $m$ is the band index.
	This is a multiparticle generalization of the conventional single-particle energy band. 

	If an initial state uniformly occupies the $m$th center-of-mass energy band,
	the mean position of particles is shifted by multiple unit cells in one pumping cycle $T$, 
	\begin{equation}
		\langle \delta x\rangle =\langle x (T) \rangle-\langle x (0) \rangle= C_m d.
	\end{equation}
	Here, $C_m$ is the Chern number of the $m$th center-of-mass energy band, which is defined as
	\begin{equation}
		C_m=\frac{i}{2\pi} \int_0^{2\pi/d} dK  \int_0^{T} dt 
		( \langle \partial_t\Psi_{m,K}|\partial_K \Psi_{m,K}\rangle-\langle \partial_K\Psi_{m,K}|\partial_t \Psi_{m,K}\rangle).
	\end{equation}
	
	Without loss of generality, we consider topological pumping via the interacting Rice-Mele model described by Equation 1 in the main text.
	Depending on the ratio between the interaction $U$ and the potential well bias $2\Delta$ of the neighboring lattices, there are two major different topological pumping phenomena in the two-body limit.
	When $\vert U\vert\gg \vert 2\Delta \vert$ and $\vert U\vert\gg~J,~\delta$, 
	two particles at the same site will form bound states, which can be effectively viewed as a composite particle.
	In one pumping cycle, the two particles will be shifted by a unit cell as a whole.
	When $\vert U\vert$ is comparable to $\vert 2\Delta \vert$ and $\vert U\vert\gg~J,~\delta$, 
	there is negligible energy offset between a bound state and two particles at nearest-neighboring sites, and resonant coupling between the two states can occur.
	In one pumping cycle, the resonant tunnelings happen four times and the two particles will be shifted one by one to a unit cell.
	In the following, we give a detailed theoretical analysis of topological pumping of bound states and topologically resonant tunnelings.
	
	\subsection{Energy band for topological pumping of bound states}
\begin{figure}[!htbp]
		\centering
		\includegraphics[width=0.8\textwidth]{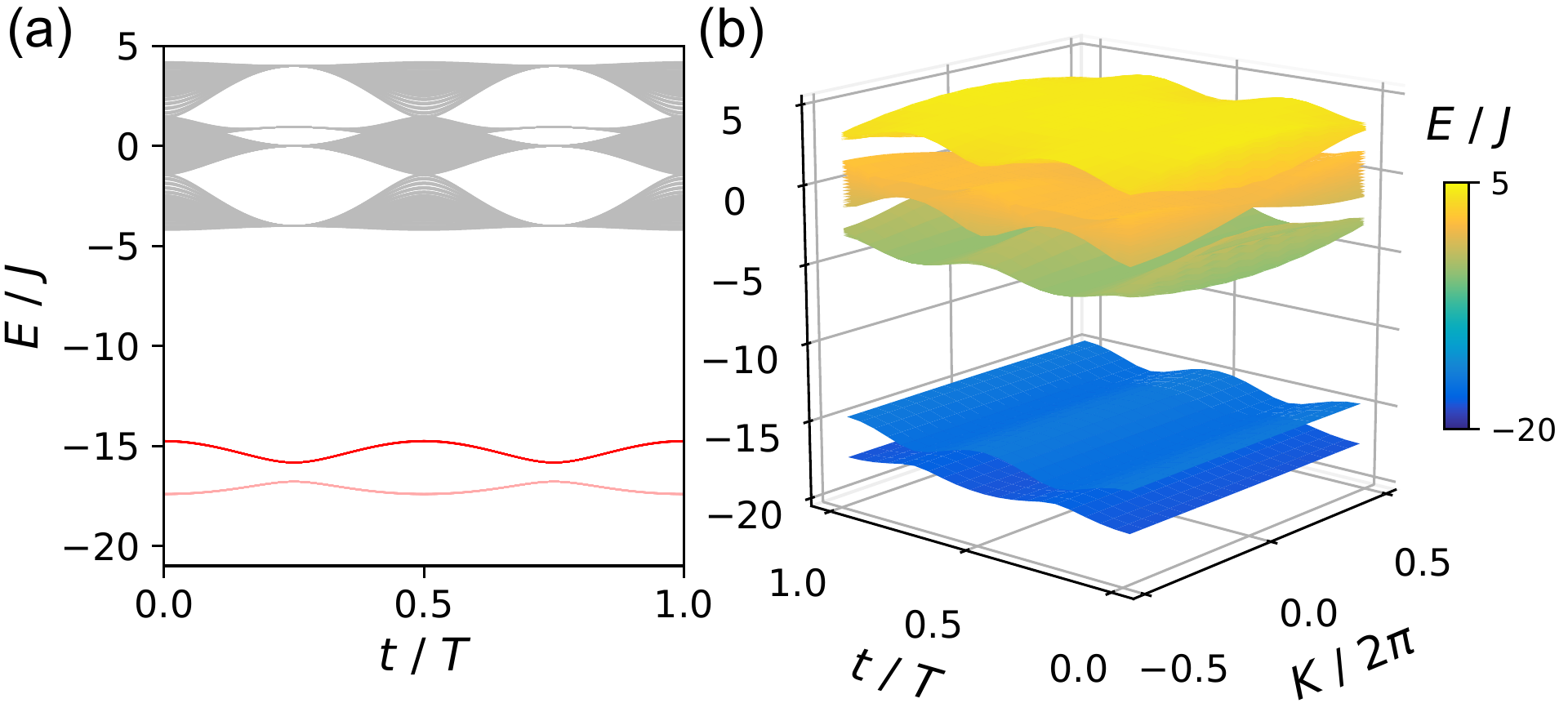}\\
		\caption{Center-of-mass energy band for topological pumping of bound states.
			(a)
			Energy spectrum as a function of time. 
			(b)
			Center-of-mass energy band as functions of time and center-of-mass momentum. Here, the two-particle bound-state pumping is performed with the parameters of $\Delta_0/2\pi = 8$~MHz, $\delta_0/2\pi = J/2\pi = 12$~MHz and a pumping period of $T=0.4~\mu$s.
		}\label{BoundStateBand}
	\end{figure}

	We calculate the center-of-mass energy band for the case of topological pumping of bound states; see \rfig{BoundStateBand}.
	The red and pink lines in \rfig{BoundStateBand}(a) indicate the energy bands for the bound states in which two particles are almost at the same site, while the grey lines indicate the energy bands for the scattering states in which two particles are almost independent.
	In the $(K,t)$ space (\rfig{BoundStateBand}(b)), the Chern numbers of the center-of-mass energy bands for bound states are $\pm 1$.
	If we prepare the initial state as two particles at the same odd (even) site, the initial state almost uniformly occupies the pink (red) energy spectrum.
	The two particles as a whole will be moved forward (backward) by a unit cell, corresponding to the Chern number $+1$ $(-1)$.
	The experimental result on topological pumping of bound states in the main text (Figure 2) is consistent with this theoretical expectation.

	\subsection{Energy band for topologically resonant tunnelings}
	\begin{figure}[!htbp]
		\centering
		\includegraphics[width=0.8\textwidth]{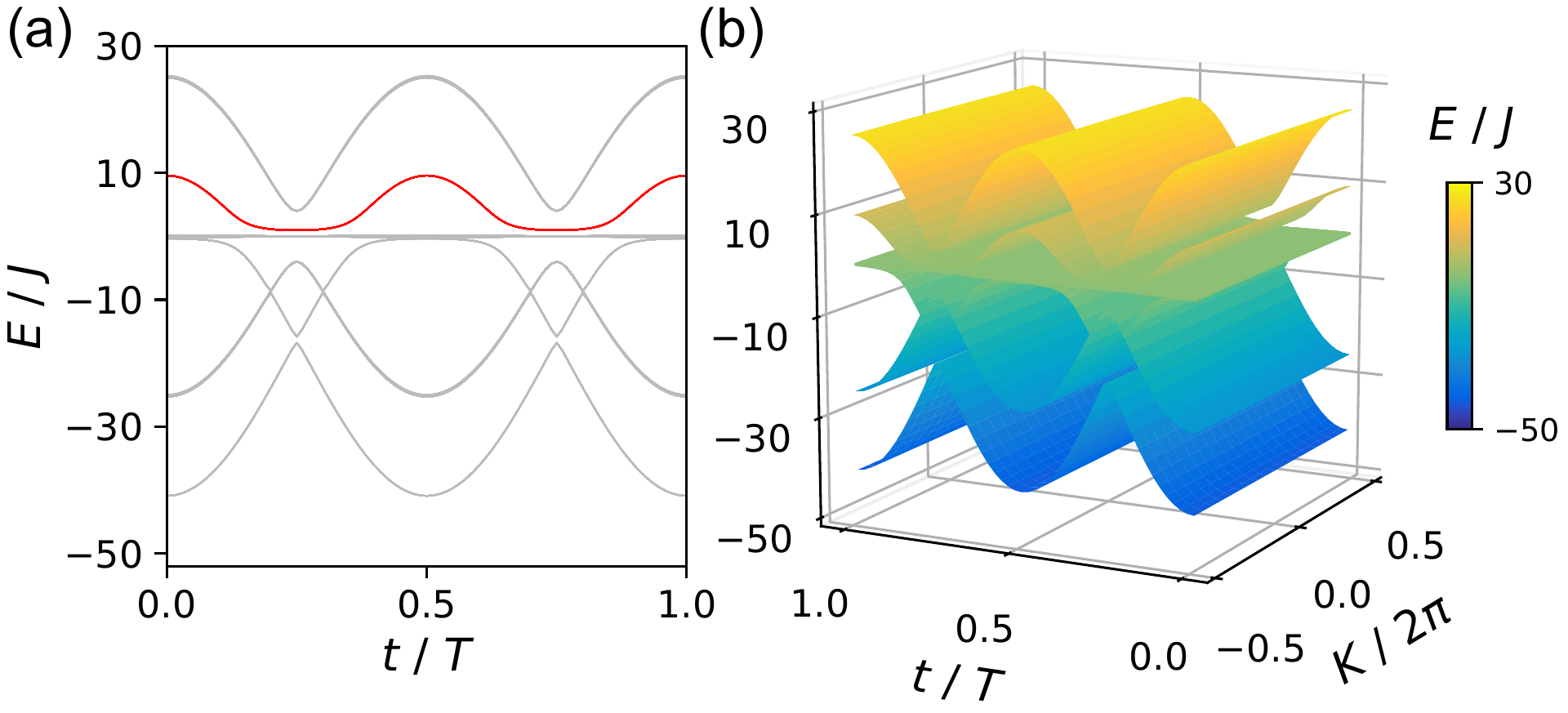}\\
		\caption{Center-of-mass energy band for topologically resonant tunnelings. 
			(a)
			Energy spectrum as a function of time.
			(b)
			Center-of-mass energy band as functions of time and center-of-mass momentum. The parameters are set as $\Delta_0/2\pi = 150$~MHz, $\delta_0/2\pi = J/2\pi = 12$~MHz and a pumping period of $T=0.4~\mu$s.
		}\label{ResonantTunnelingBand}
	\end{figure}
	The center-of-mass energy band for the case of topologically resonant tunnelings is also calculated, as shown in  \rfig{ResonantTunnelingBand}.
	Compared with the case of topological pumping of bound states, the energy spectrum becomes more complex.
	When the energy is around $0$, there is an isolated energy band (red lines in \rfig{ResonantTunnelingBand}(a)) that can explain topologically resonant tunnelings.
	Around the energy-avoided crossing, the states $\vert 2_j\rangle \vert 0_{j+1}\rangle$ and $\vert 1_j\rangle \vert 1_{j+1}\rangle$ are strongly coupled, leading to the resonant tunneling between them.
	There are four energy-avoided crossing points, indicating that resonant tunnelings occur four times.
	In \rfig{ResonantTunnelingBand}(b), the Chern number of the isolated band for topologically resonant tunnelings is $+1$.
	If we prepare the initial state as two particles at the same site, such an initial state almost uniformly occupies the isolated band around $0$, the subsequent four resonant tunnelings give rise to quantized forward shift of two particles one by the other.

	We emphasize that the topologically resonant tunneling is interaction-induced topological pumping that has no linear or noninteracting counterpart.
	If we start from the initial Hamiltonian with no energy bias between adjacent sites, the initial state of two particles $\vert 1_j\rangle \vert 1_{j+1}\rangle$ can be expanded as a superposition of two symmetric and asymmetric states, 
	\begin{equation}
		\vert 1_{j}\rangle \vert 1_{j+1}\rangle=
		\frac{1}{2\sqrt{2}}\left[
		(\vert 1_{j}\rangle + \vert 1_{j+1}\rangle)\otimes
		(\vert 1_{j}\rangle + \vert 1_{j+1}\rangle)
		-
		(\vert 1_{j}\rangle-\vert 1_{j+1}\rangle)\otimes
		(\vert 1_{j}\rangle-\vert 1_{j+1}\rangle)
		\right].
	\end{equation}
	Without inter-particle interaction, the symmetric and asymmetric states uniformly occupy the single-particle lower and upper bands, which have opposite Chern numbers $\pm 1$, respectively.
	The symmetric and asymmetric states are respectively shifted forward and backward by one unit cell, and the center-of-mass positions cancel with each other, with no shift over a pumping cycle.

	\subsection{Interaction-induced asymmetric topological pumping of edge states}
	\begin{figure}[!htbp]
		\centering
		\includegraphics[width=0.8\textwidth]{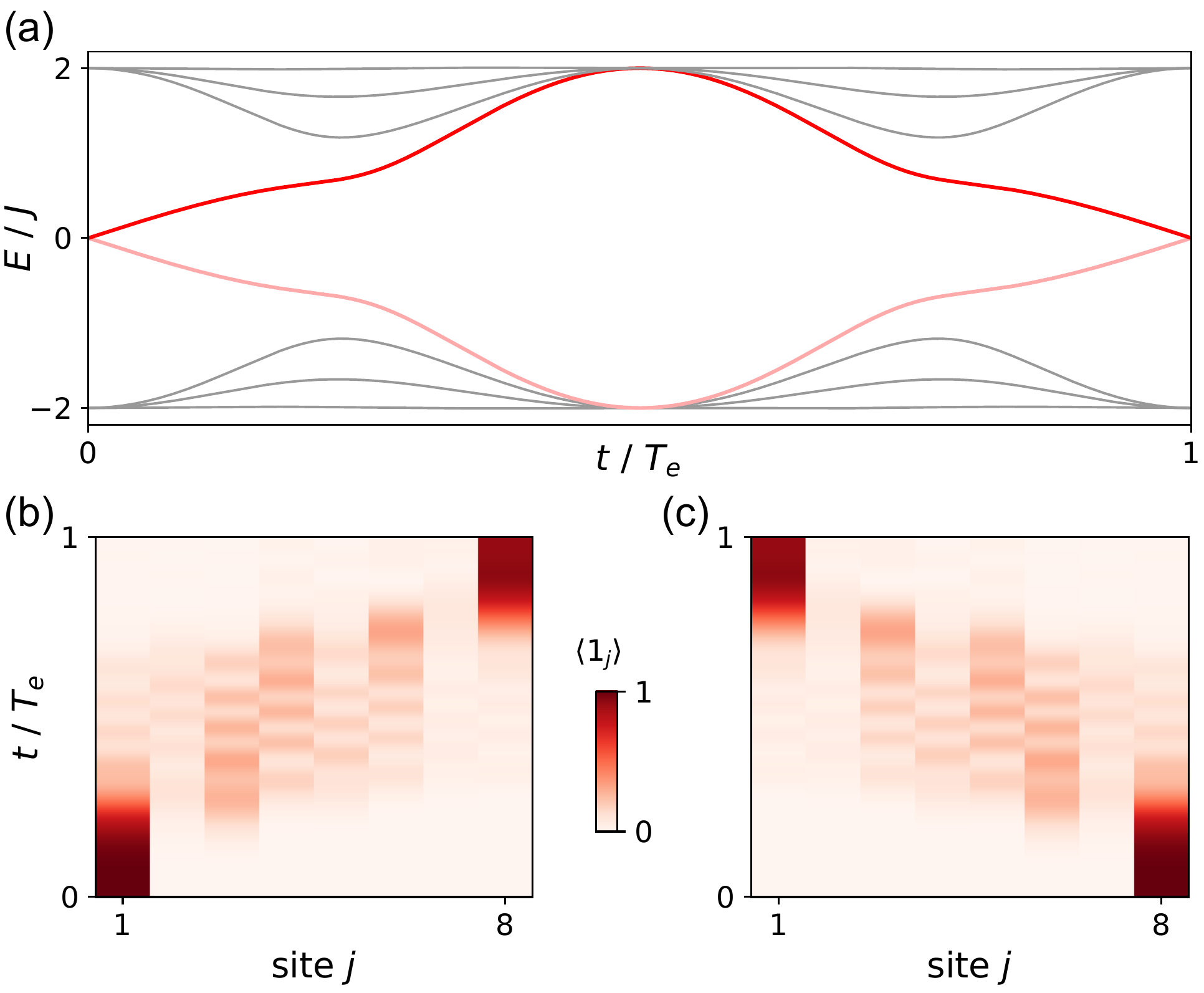}\\
		\caption{
			(a)
			Symmetric energy spectrum of single-particle Rice-Mele model.
			(b)
			Topological pumping of a particle from left to right edges.
			(c)
			Topological pumping of a particle from right to left edges. The parameters are chosen as $\delta_0/2\pi=J/2\pi=0.75~\mathrm{MHz}$, $\Delta_0/2\pi=0.5~\mathrm{MHz}$, and $T_e=4~\mu\mathrm{s}$, 
			where $\Delta_0$ and $T_e$ are the same as two-particle edge state pumping in the main texts, 
			and $\delta_0$, $J$ are chosen to be comparable with the effective coupling $J_{\rm{eff}}\simeq (J+\delta)^2/U$ between $\vert 2_j\rangle$ states\cite{Ke2017} in the main texts.
		}\label{SingleSymmetry}
	\end{figure}
	
	In this section, we discuss how the inter-particle interaction affects topological pumping of edge states.
	Before taking the interaction into account, we first consider the single-particle case, in which the energy spectrum of the Rice-Mele model is symmetric with respect to zero energy; see \rfig{SingleSymmetry}(a).
	There are two topological edge modes in the bulk gap, which provide reversible channels for topological pumping of edge states from the left end to the right end or vise versa. 
	Because the energies of the two channels are symmetric about zero,
	the topological pumping from the left to the right edges is symmetric with the reversed process; see \rfigs{SingleSymmetry}(b) and (c).

	\begin{figure}[!htbp]
		\centering
		\includegraphics[width=0.95\textwidth]{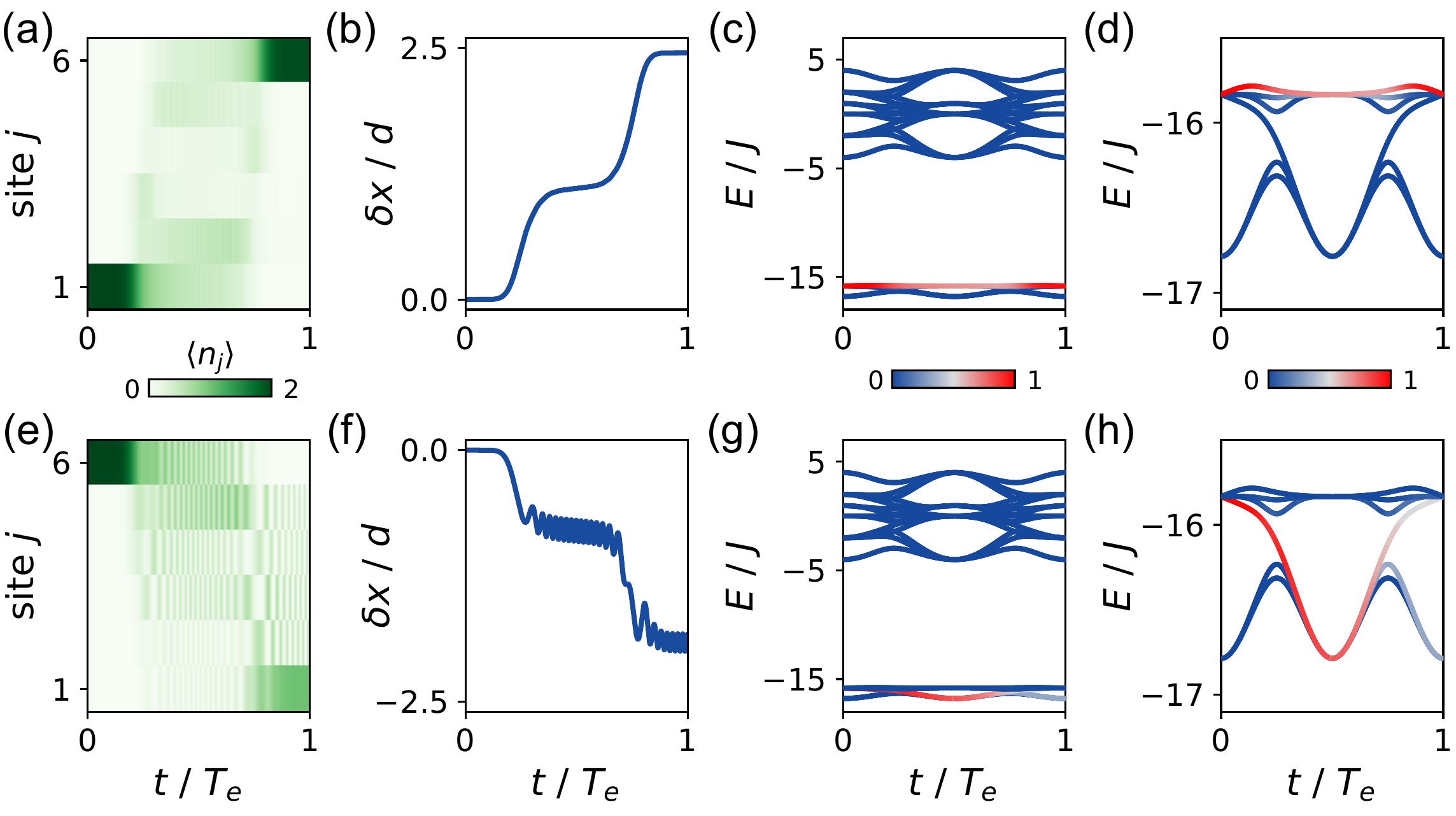}\\
		\caption{
			Top panel: topological pumping of two particles from left to right edges.
			(a)
			Density evolution in which the colors denote the density of particles $\langle n_j\rangle$.
			(b)
			Mean position shift.
			(c)
			Energy spectrum in which the colors denote the probability of the instantaneous state projected into the eigenstates.
			(d)
			The enlarged view of (c).
			Bottom panel: breakdown of topological pumping of two particles from right to left edges. (e-h) are similar to (a-d) but with different initial state. 
			The colors in (a,e) denote the density distribution, while the colors in (c,d,g,h) denote the probability of instantaneous states occupying the eigenstates.
			The parameters are chosen as $J/2\pi=12$ MHz, $\delta_0/2\pi=12$ MHz, $\Delta_0/2\pi=0.5$ MHz, $U/2\pi=-190$ MHz, $T=4~\mu$s.
		}\label{TwoAsymmetry}
	\end{figure}
	
	However, when inter-particle interaction is involved, the picture becomes totally different.
	We consider two initial states in a lattice with $6$ sites, one is two particles at the left edge site and the other is two particles at the right edge site.
	%
	We observe that both the density evolution (\rfigs{TwoAsymmetry}(a) and (e)) and center-of-mass position shift (\rfigs{TwoAsymmetry}(b) and (f)) are different in the two cases.
	This is because the inter-particle interaction breaks the symmetric energy spectrum in the single-particle case; see \rfigs{TwoAsymmetry}(c,g) and their enlarged views in \rfigs{TwoAsymmetry}(d,h).
	In the first case, the instantaneous state can follow the eigenstates at the top of the bound-state band, which can be seen as a pumping channel from the left to the right edges (see the red eigenstates in \rfig{TwoAsymmetry}(d)).
	In the latter case, the instantaneous state can follow the eigenstates in the gap between two bound-state bands, which can be viewed as a pumping channel from the right to the left edges (see the red eigenstates in \rfig{TwoAsymmetry}(h)).
	Because both the energy gaps in the two cases are smaller than those in the bulk Thouless pumping, the adiabatic conditions are more stringent than those in the bulk Thouless pumping.
	Besides, the different energy gaps in the two cases also lead to different adiabatic conditions.
	\begin{figure}[H]
		\centering
		\includegraphics[width=0.65\textwidth]{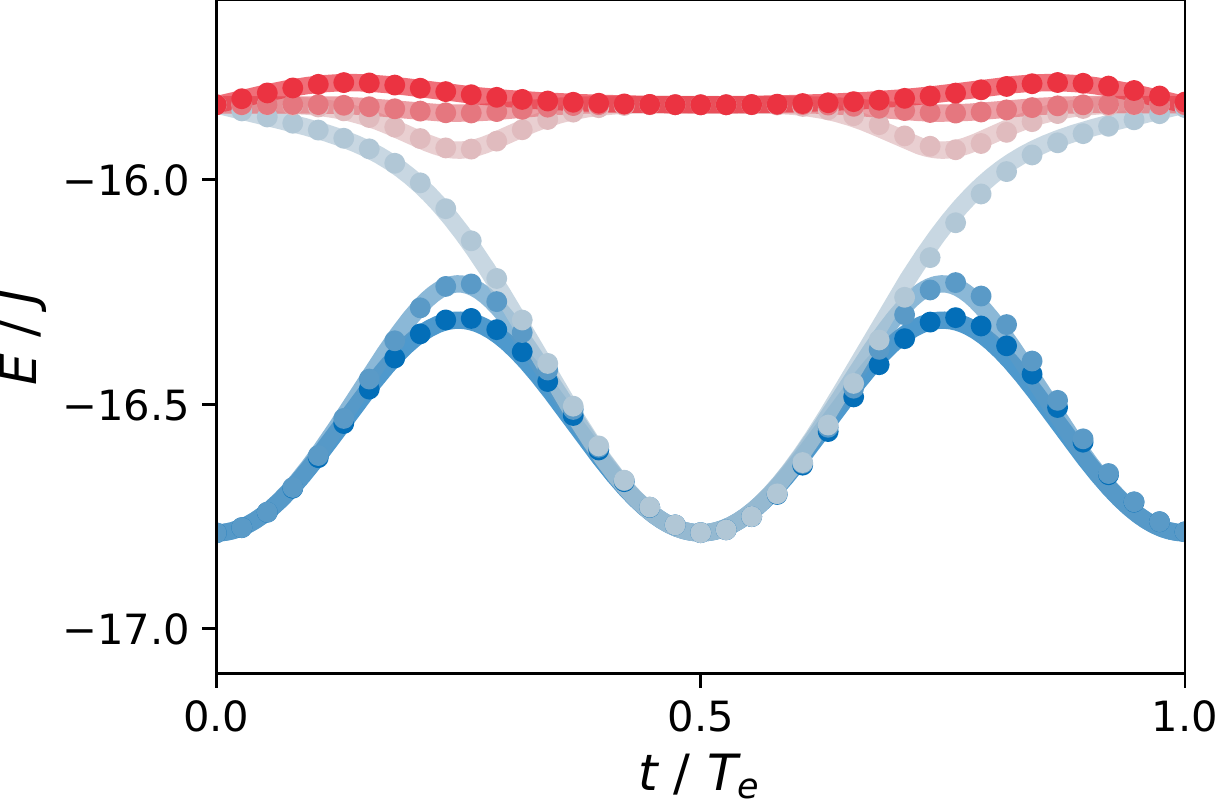}\\
		\caption{The lowest $6$ energies as functions of time. Solid lines are obtained by the original interacting Rice-Mele model and the dots are obtained by the effective Hamiltonian in the subspace of $\mathcal V$ consisting of 
			$\{\vert 2_j\rangle\}$ and $\{\vert 1_j\rangle \vert 1_{j+1}\rangle\}$. 
			The parameters are the same as those in \rfig{TwoAsymmetry}. 
		}\label{SpectrumEffect}
	\end{figure}
	
	To better understand the physical picture of the asymmetric spectrum in \rfigs{TwoAsymmetry}(d) and (h), we can approximate the subspace $\mathcal V$ consisting of $\{\vert 2_j\rangle\}$ and $\{\vert 1_j\rangle \vert 1_{j+1}\rangle\}$.
	In the case of $|U|\gg J, \delta,\Delta$, 
	the states of two neighboring particles $\{\vert 1_j\rangle \vert 1_{j+1}\rangle\}$ form intermediate states which can be adiabatically eliminated, 
	and we can obtain an effective Hamiltonian for the bound states $\{\vert 2_j\rangle\}$, which turns out to be similar to single-particle Rice-Mele model~\cite{Ke2017}.
	However, in the present case where $|U|$ is modest, the subspace of $\{\vert 1_j\rangle \vert 1_{j+1}\rangle\}$ cannot be neglected.
	In the subspace $\mathcal V$, the matrix of the effective Hamiltonian inherits the original interacting Rice-Mele model.
	To be explicit, the coupling strength between $\vert 2_j\rangle$ and $\vert 1_j\rangle\vert 1_{j+1}\rangle$ is $\sqrt{2}[-J+(-1)^{j}\delta]$, 
	the onsite energies of $\vert 2_j\rangle$ and $\vert 1_j\rangle\vert 1_{j+1}\rangle$ are $2(-1)^j\Delta+U$, and $0$, respectively.
	By diagonalizing the effective Hamiltonian, we find that the lowest $6$ energies (see \rfig{SpectrumEffect}) are consistent with the spectrum in  \rfigs{TwoAsymmetry}(d) and (h).
	Hence, the modest interaction plays a crucial role in the asymmetric spectrum, which cannot be reduced to the simple single-particle Rice-Mele model.

	%
	
	\subsection{Mapping from Topological pumping to integer quantum Hall effect}
	For a periodically driven Hamiltonian $H(t)=H(t+T)$, in a Hilbert space formed by Fock states $\{\vert n_1,n_2,...,n_N \rangle\}$, 
	the quasieigenstates can be written as 
	\begin{equation} \label{quasistate}
		\vert \Psi(t)\rangle=
		e^{-i \varepsilon_{\alpha} t} 
		\sum\limits_{m,n_1,n_2,...,n_N}e^{i m\omega t}C_{m,n_1,n_2,...,n_N}^{\alpha}
		\vert n_1,n_2,...,n_N\rangle,
	\end{equation}
	where $C_{m,n_1,n_2,...,n_N}^\alpha$ are the amplitudes of the $\alpha$th eigenstate of the time-independent Floquet operator $\mathcal H$, with elements
	\begin{equation}
		\mathcal{H}_{m,n_1,n_2,...,n_N;m',' n_1',n_2',...,n_N'}=
		\frac{1}{T}\int_0^T  \langle n_1,n_2,...,n_N|e^{-im\omega t} [H(t)-i\partial_t] e^{im'\omega t} \vert n_1',n_2',...,n_N'\rangle dt.\label{HM}
	\end{equation}
	The term of derivative of time contributes to a tilting potential $(m\omega)$ along the Floquet dimension.
	In the case of a single particle, the states can be alternatively represented by the position of a particle $\{|j\rangle\}$, and the amplitude $C_{m,j}^{\alpha}$ can be obtained by diagonalizing the Floquet Hamiltonian with elements~\cite{Platero2013},
	\begin{equation}
		\mathcal H _{m,j;m',j'}=\frac{1}{T}\int_0^T  \langle j|e^{-im\omega t} [H(t)-i\partial_t] e^{im'\omega t}|j'\rangle dt.\label{H1}
	\end{equation}
	
	Before proceeding to the relation between Thouless pumping and integer quantum Hall effect,
	we make some remark on the general mapping from one dimension to two dimensions in the single-particle and multi-particle cases based on Eqs.~\ref{HM} and \ref{H1}.
	In the single-particle case, a particle in a one-dimensional periodically driven system can be mapped to a two-dimensional static system with a tilting potential along the Floquet dimension.
	In the multi-particle case with particle conservation, in a given Floquet index $m$, %
	the particle number is fixed.
	When mapping to synthetic two dimensions, all particles as a whole are transferred from the $m$th to $m'$th Floquet indices.
	The tilting energy $m\omega$ is also independent of the particle density. 
	It is totally different from the two-dimensional real-space lattice, where particles could independently hop between different lattices, and the total energy from the tilting potential depends on the particle density.
	%
	%
	
\begin{figure}[!htbp]
		\centering
		\includegraphics[width=0.5\textwidth]{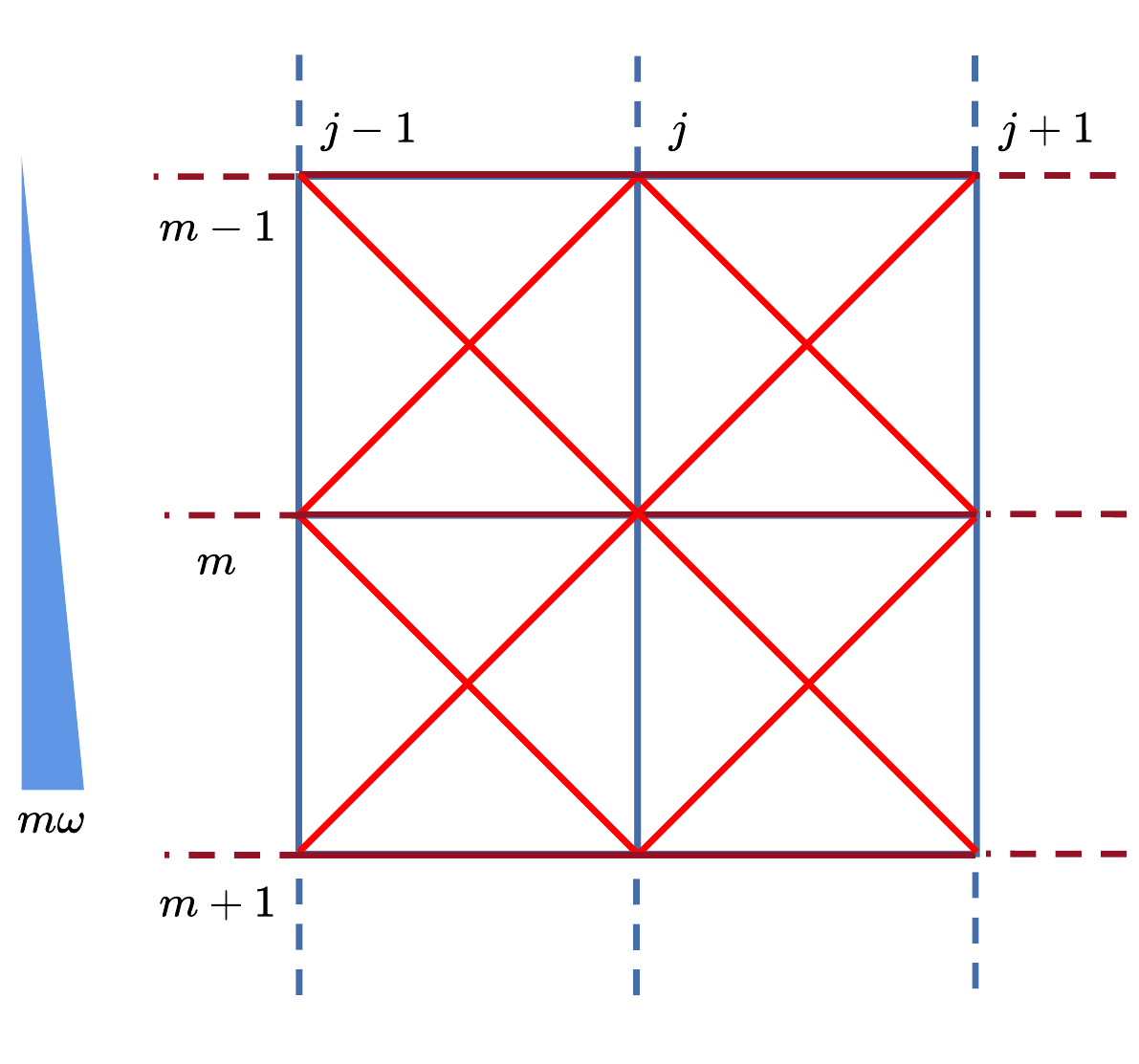}\\
		\caption{{The mapping of Rice-Mele model to a two-dimensional synthetic lattice for integer quantum Hall effect. Apart from Harper-type model, There is an additional tilting potential along the synthetic direction. }
		}\label{Map}
	\end{figure}

	Taking the Rice-Mele model as an example, in which $\delta(t)=\delta_0 \sin(\omega t+\phi_0)$ and $\Delta(t)=\Delta_0 \cos(\omega t+\phi_0)$ with modulation phase $\phi_0$,  we will illuminate the one-to-one correspondence between Thouless pumping and the integer quantum Hall effect.
	In the single particle case, the elements of the Floquet operator are given by
	\begin{eqnarray} \label{FloquetOne}
		\mathcal H_{m,j,m',j'}&=&\delta_{m,m'}\left[m\omega -J(\delta_{j,j'+1}+\delta_{j,j'-1})\right],  \\
		&+&\delta_{m,m'+1}\left[\frac{\delta_0 e^{i\phi_0}(-1)^{j}}{2i}\delta_{j,j'-1}+\frac{\delta_0 e^{i\phi_0}(-1)^{j'}}{2i}\delta_{j,j'+1}+\frac{\Delta_0 e^{i\phi_0}(-1)^{j'}}{2}\delta_{j,j'}\right] \nonumber \\
		&+&\delta_{m,m'-1}\left[-\frac{\delta_0 e^{-i\phi_0}(-1)^{j}}{2i}\delta_{j,j'-1}-\frac{\delta_0 e^{-i\phi_0}(-1)^{j'}}{2i}\delta_{j,j'+1}+\frac{\Delta_0 e^{-i\phi_0}(-1)^{j'}}{2}\delta_{j,j'}\right], \nonumber 
	\end{eqnarray}
	where $\delta_{m,m'}$ is the Kronecker delta function. To better understand the above equation, we show the connection of the synthetic two-dimensional lattice; see \rfig{Map}.
	We can find that there is tilting potential along the synthetic lattice direction and three kinds of hopping processes: nearest-neighboring hopping along the real lattice direction, nearest-neighboring hopping along the synthetic lattice direction due to time modulation of onsite energy, and next-nearest neighboring hopping due to time modulation of hopping strength.
	We first neglect the tilting potential and find that the two-dimensional synthetic lattice has translational symmetry. 
	There are two lattice sites in a unit cell, indicating that there are two energy bands in the momentum space.
	The energy bands can be obtained by applying the Bloch theorem, with which the Bloch functions are given by
	\begin{equation} \label{BlochFun}
		\begin{aligned} 
			& \psi_{k_x,k_y}(2j-1,m)= A e^{-ik_y m} e^{-ik_x (2j-1)}, \\
			& \psi_{k_x,k_y}(2j,m)=B e^{-ik_y m} e^{-ik_x 2j},
		\end{aligned}
	\end{equation}
	where $k_{x(y)}$ is the quasimomentum in the $x(y)$ direction.
	By solving the Schr\"odinger equation $\mathcal H \psi_{k_x,k_y}=E\psi_{k_x,k_y}$, we can find that the coefficients $(A, B)$ can be obtained by diagonalizing the following Hamiltonian,
	\begin{equation}
		H_{k_x,k_y}=\left[\begin{matrix}
			-\Delta_0\cos(k_y+\phi_0) & 2[J\cos(k_x)+i\delta_0 \sin(k_y+\phi_0)\sin(k_x)] \\
			2[J\cos(k_x)-i\delta_0 \sin(k_y+\phi_0)\sin(k_x)] & \Delta_0\cos(k_y+\phi_0) \\
		\end{matrix}
		\right]
	\end{equation}
	This equation is equivalent to the Hamiltonian in the $(k,t)$ space~\cite{Ke2020}, if we replace $k_y$ with $\omega t$.
	It means that the topological phases between the two-dimensional synthetic lattice and the one-dimensional periodically modulated lattice are the same.  
	Now, we can discuss the relation between Thouless pumping and integer quantum Hall effect.
	In both cases, we prepare an initial Wannier state which uniformly occupies the Bloch state in a given band along the $k_x$ direction.
	In the one-dimensional periodically modulated lattice, the mean position shift in a pumping cycle is equal to a unit cell.
	While in the two-dimensional synthetic lattice, due to the tilting potential, the particle will undergo a Bloch oscillation along the synthetic direction, and the mean position shift along the real-space direction in a period of Bloch oscillation is also equal to a unit cell.
	The mean position shift along the real-space direction is the same in both cases.
	That is why the Thouless pumping can be viewed as a dynamical version of integer quantum Hall effect.
	
	In the multi-particle cases,  the mapping from Thouless pumping to two-dimensional integer quantum Hall effect is somehow tricky.
	We cannot simply add interacting particles in the two-dimensional synthetic lattice subjected to tilting potential along the synthetic direction. 
	This is because we need to consider the restriction that the particle number is conserved for a given Floquet index $m$.
	It means that all the particles jump between subspaces with different Floquet indices as a whole.
	The particles independently tunnel along the real-space direction but behave as bound states along the synthetic direction.
	This is in stark contrast to the case of interacting particles in a real two-dimensional lattice.

	\subsection{Numerical simulation of qubit dynamics}
	We give the numerical simulation of qubit dynamics using the Lindblad master equation with the QuTiP python library\cite{Johansson2012}.
	The full Hamiltonian is given as 
	\begin{equation}\label{hami_simu_qubit}
		H/\hbar =
		\sum_{j=1}^{N-1} \left[[-J + (-1)^j \delta] {a}_j ^\dag {a}_{j+1} + \textrm{H.c.}\right] + 
		\sum_{j=1}^{N} \left[(-1)^j \Delta {a}^\dagger_j {a}_j + 
		\frac{U}{2} {a}^\dagger_j {a}^\dagger_j {a}_j{a}_j \right],
	\end{equation}
	where
	\begin{equation}
		\begin{aligned} 
			{a}_j &= I\otimes I \otimes \cdots {a} \otimes I \cdots, \\
			{a}^\dagger_j &= I\otimes I \otimes \cdots {a}^\dagger \otimes I \cdots , \\
		\end{aligned}
	\end{equation}
	$\otimes$ is tensor product.
	The lowest three energy levels $\vert n\rangle$ ($n=0,1,2$) of our superconducting qubits are considered, 
	$\vert 0\rangle = (1,0,0)^T$, 
	$\vert 1\rangle = (0,1,0)^T$,
	$\vert 2\rangle = (0,0,1)^T$,
	which give the annihilation and creation operators in a matrix form
	\begin{equation}
		{a} = 
		\begin{pmatrix}
			0 & 1 & 0\\
			0 & 0 & \sqrt{2}\\
			0 & 0 & 0\\
		\end{pmatrix}, 
		\quad
		{a}^\dagger = 
		\begin{pmatrix}
			0 & 0 & 0\\
			1 & 0 & 0\\
			0 & \sqrt{2} & 0\\
		\end{pmatrix}.
	\end{equation}
	As the qubit decoherence parameters depend on the qubit fequencies, it is difficult to accurately model pumping dynamics with time-varying decoherence. For simplicity, the numerical calculation results in the main text are obtained without decoherence.
	
\begin{figure}[!htbp]
		\centering
		\includegraphics[width=0.8\textwidth]{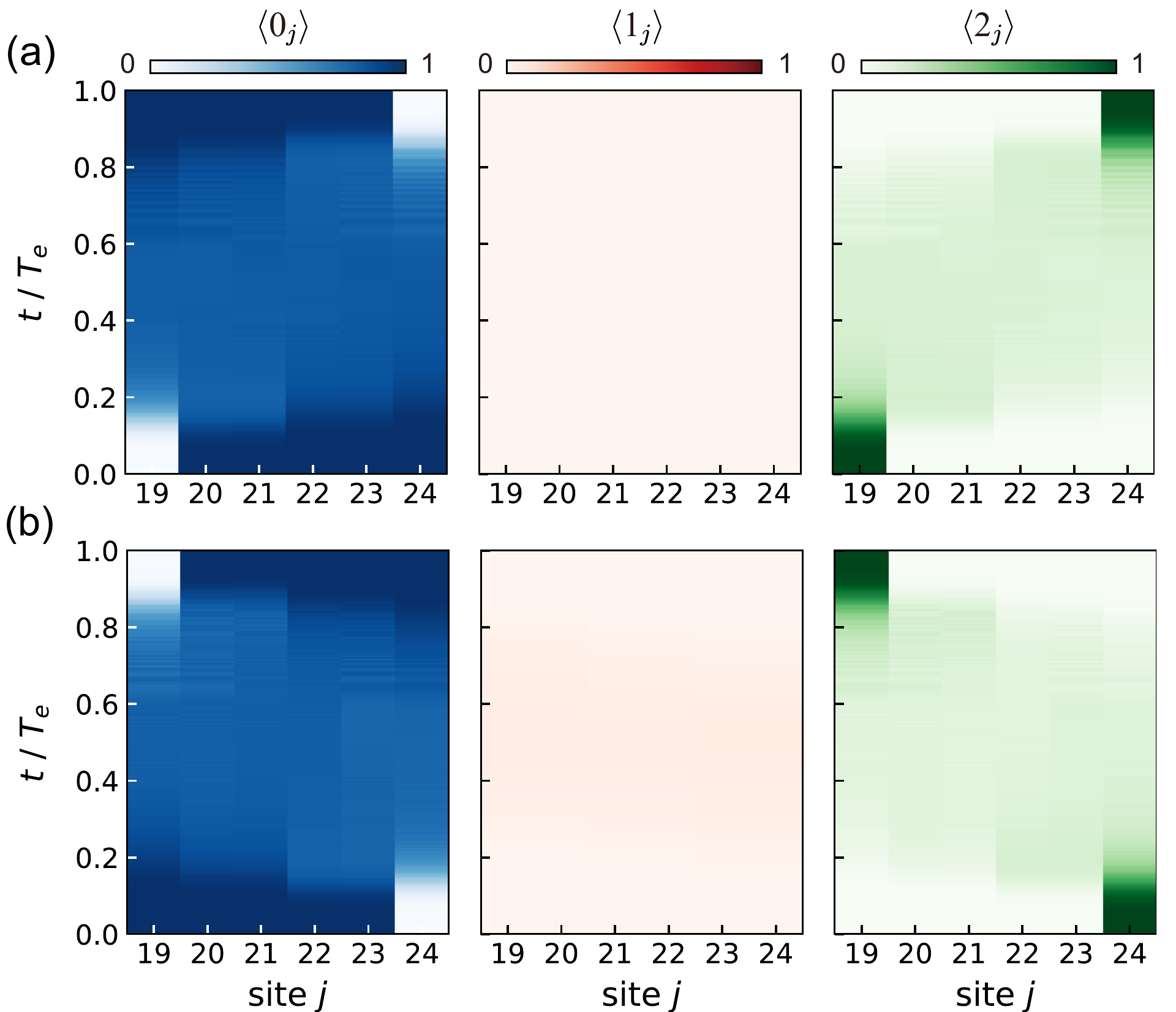}
		\caption{\label{supply_edge2evo_40us} 
			Numerical simulation of the 
			left (a) and right (b) edge state pumping for $T_e=40~\mu$s, 
			where $\Delta_0/2\pi=0.5~\mathrm{MHz}$, $\delta_0/2\pi=J/2\pi=25~\mathrm{MHz}$.
		}
	\end{figure}

	\rfig{supply_edge2evo_40us} gives the numerical results of topological pumping of two-particle edge states  with the period $T_e=40~\mu\textrm{s}$, 
	where the non-adiabatic effects of pumping from the right edge state are suppressed and the dynamics become similar to the reversal pumping from the left edge state, as compared to the case of short $T_e=4~\mu\textrm{s}$ in Figure 4 in the main text.

	\clearpage
	
	\section{Robustness of topological pumping}
	We further explore the robustness of single-particle topological pumping to different sources of imperfection, including the pumping period, on-site interaction disorder, on-site energy offset and disorder.
	\subsection{Adiabaticity versus decoherence}
\begin{figure}[!htbp]
		\centering
		\includegraphics[width=0.65\textwidth]{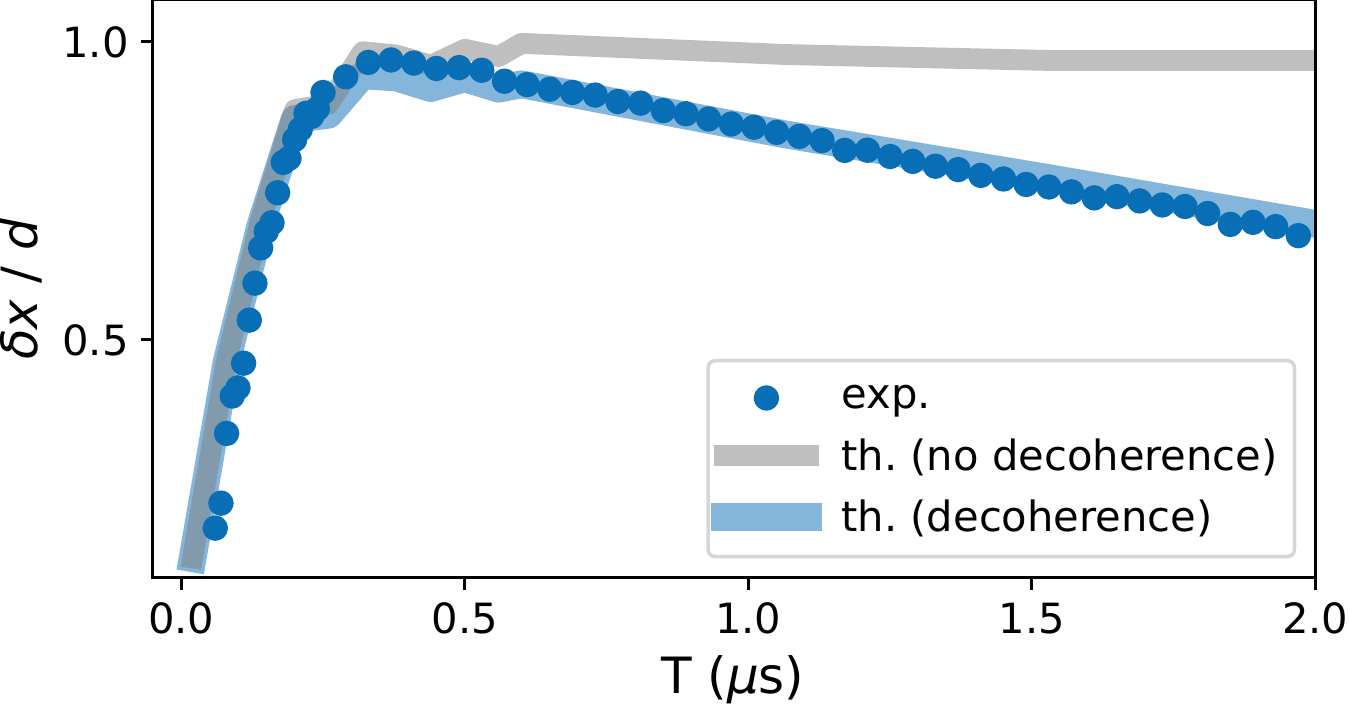}
		\caption{\label{supply_pump_changeT} 
			Center of mass displacement $\delta x= d/2\sum j n_j / \sum n_j$ for single-particle pumping with different pumping period $T$. 
			The blue dots are experimental data, the blue and grey curves are numerical simulations with and without decoherence, respectively.
			The effective energy relaxation time $T_1^{\rm{eff}}=25~\mu\rm{s}$ and pure dephasing time $T_{\phi}^{\rm{eff}}=1~\mu\rm{s}$ are considered for the blue curve, 
			which are contributed from the qubit dephasing away from the sweet spot and the coupling between the qubit and the less coherent coupler\cite{Xu2020a}.
		}
	\end{figure}

	\rfig{supply_pump_changeT} shows single-particle topological pumping with different pumping period $T$, where the non-adiabatic effects for a short $T$ and decoherence for a long $T$ contribute to the deviations from ideal pumping, respectively.
	Since the energy relaxation and pure dephasing times during the qubit and coupler frequency modulation are dynamical and much shorter than the case at the idle point\cite{Xu2020a}, we consider a static effective energy relaxation and pure dephasing time in the numerical simulation of master equation which fits the experimental results well with long pumping periods.
	
	\subsection{On-site interaction disorder}
\begin{figure}[!htbp]
		\centering
		\includegraphics[width=0.8\textwidth]{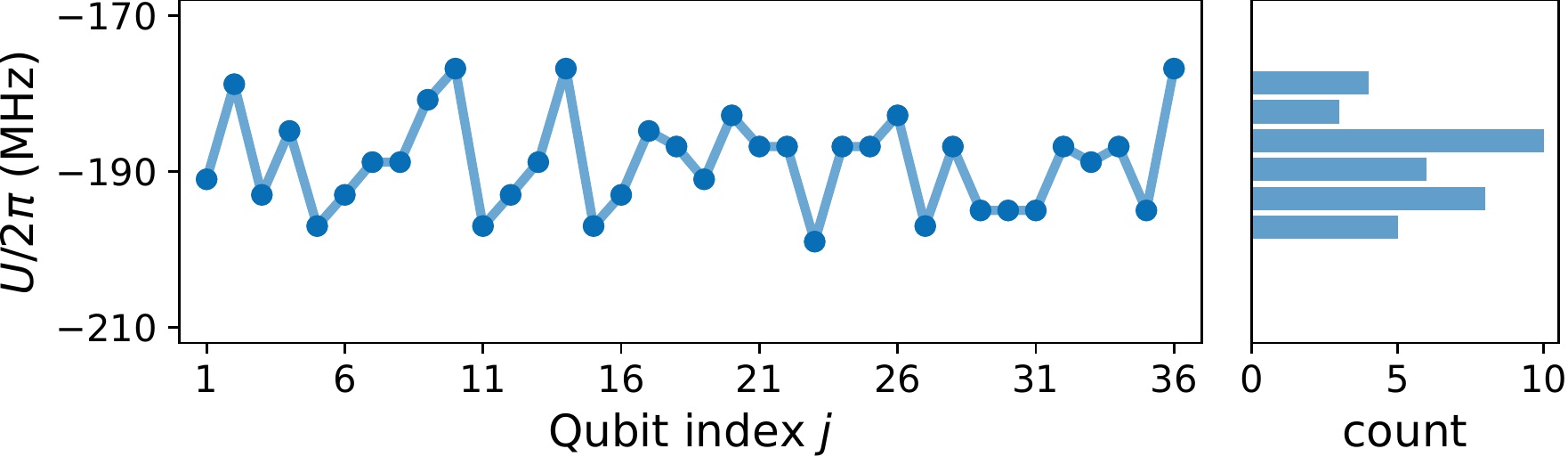}
		\caption{\label{qubit_U_count} {Interaction disorder.}
		}
	\end{figure}

	In this experiment, the strength of the on-site interaction varies for different sites due to the imperfection of the sample, as shown in \rfig{qubit_U_count}.
	
\begin{figure}[!htbp]
		\centering
		\includegraphics[width=0.8\textwidth]{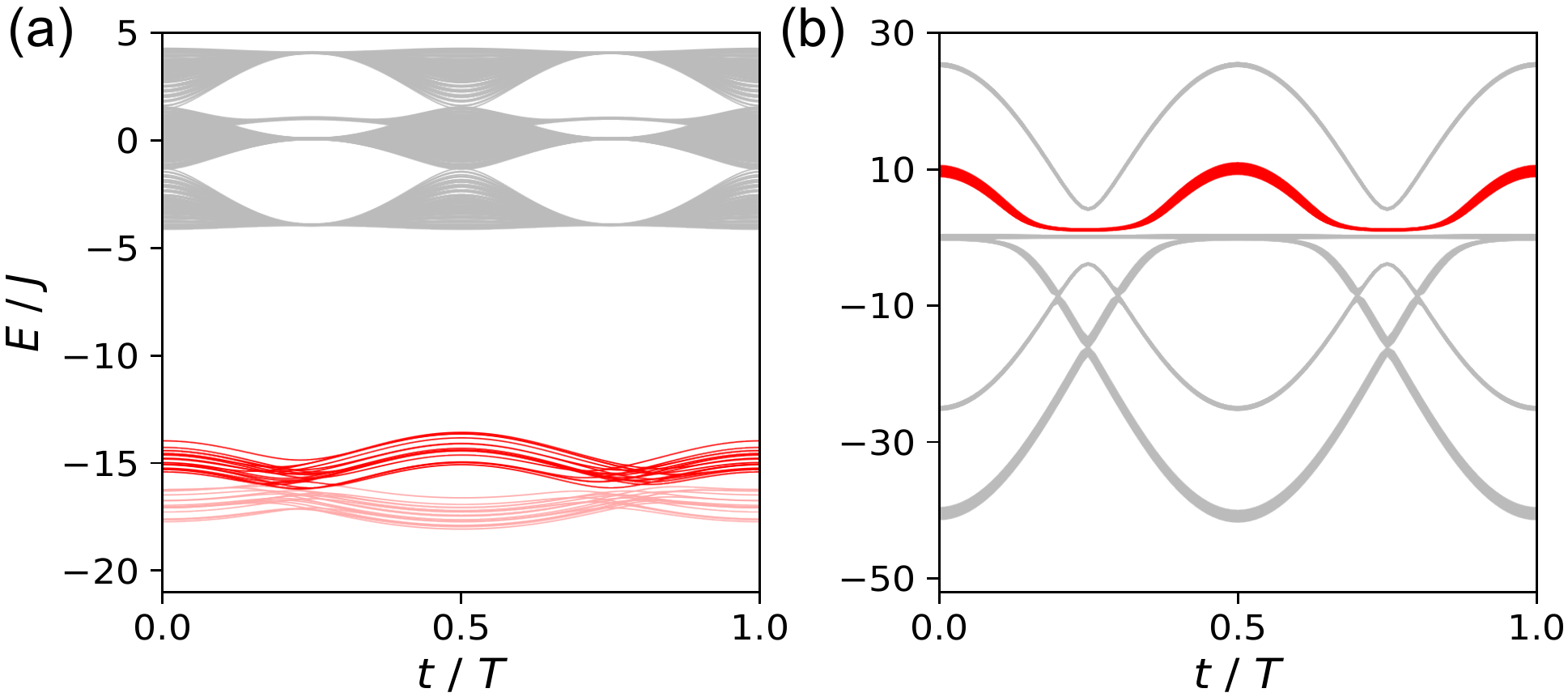}
		\caption{\label{SM_Disordered_Spectrum} {Energy spectrum in the presence of disordered interaction.}
			(a) and (b) show the energy spectrum during the pumping of bound states and topologically resonant tunneling respectively.
			The bound state bands we focus on are marked by red and pink colors.
			The parameters are the same as \rfig{BoundStateBand} and \rfig{ResonantTunnelingBand} except for the disordered interaction strength given in \rfig{qubit_U_count}.
		}
	\end{figure}

\begin{figure}[!htbp]
		\centering
		\includegraphics[width=0.6\textwidth]{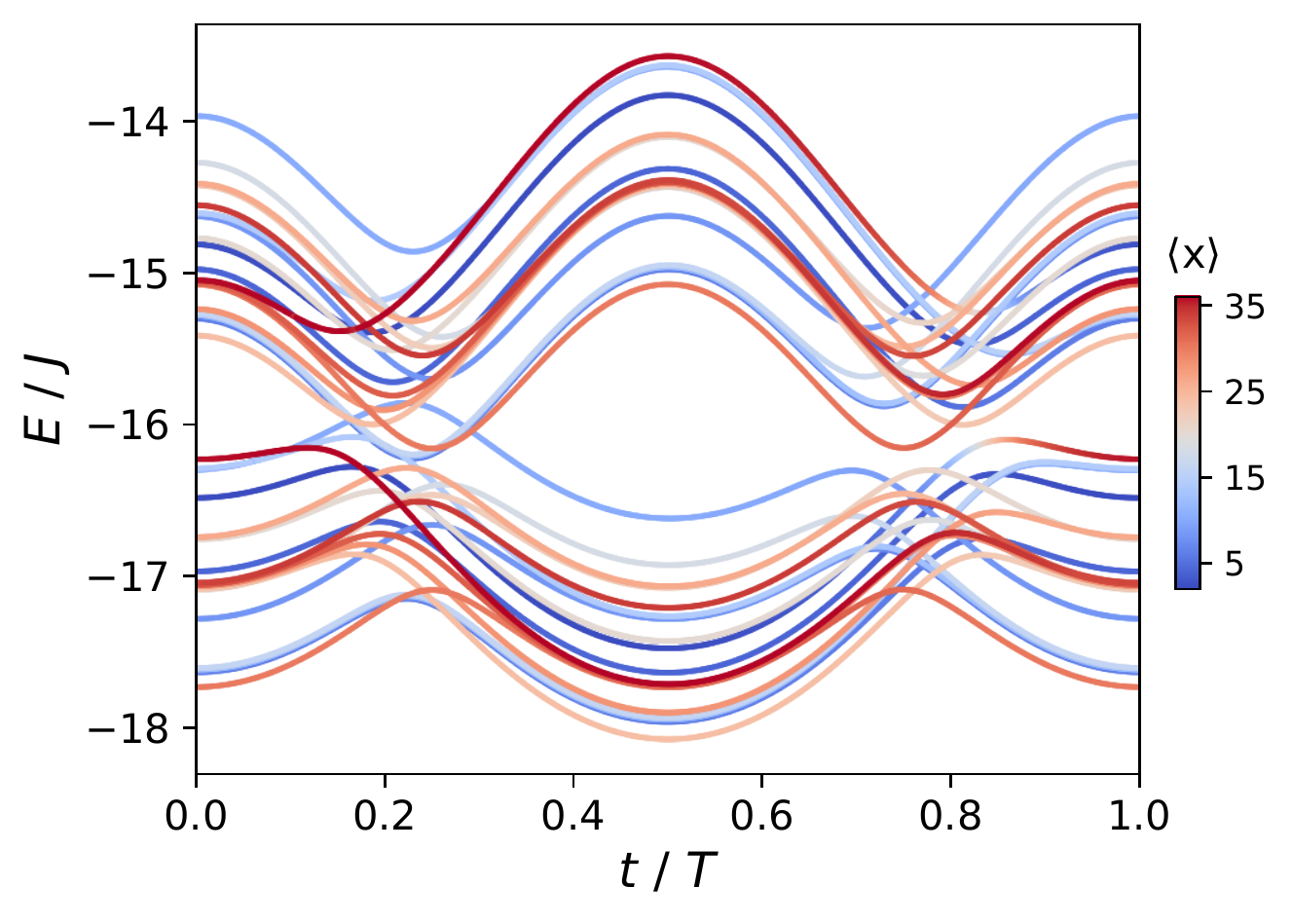}
		\caption{\label{SM_Disordered_Spectrum_enlarged} {Enlarged area of the energy spectrum.}
			We mark each eigenenergy by the average position $\langle x \rangle = \langle \psi|\hat x |\psi \rangle$.
			The parameters are the same as \rfig{BoundStateBand}.
		}
	\end{figure}

	It is of importance to study the  effect of disordered interaction on the topological pumping of bound states.
	In 1D systems, the eigenstates become localized in the presence of disordered tunneling strength or on-site potential according to the Anderson's theory \citep{PhysRev.109.1492}.
	Similarly, the disordered on-site interaction strength will lead to a localization effect on the bound states.
	We show the energy spectrum during the pumping process by considering the disordered on-site interaction in \rfig{SM_Disordered_Spectrum}.
	Notably, the bound state bands in \rfig{SM_Disordered_Spectrum}(a) are mixed near $t=T/4$ and $t=3T/4$, and the band corresponding to the resonant tunneling in \rfig{SM_Disordered_Spectrum}(b) stays gapped during the whole pumping process.
	It seems that the topological pumping of bound states will be broken in \rfig{SM_Disordered_Spectrum}(a), since the energy gap vanishes and therefore the adiabatic condition is broken.
	However, due to the disordered interaction, eigenstates are localized at different sites, and they are not coupled unless the two wavefunctions are overlapped.
	To demonstrate this effect, we calculate the average position of the eigenstate and mark them in the energy spectrum, as shown in \rfig{SM_Disordered_Spectrum_enlarged}.
	It can be seen that the eigenstates located at different sites are independent during the pumping process because they do not overlap.
	Meanwhile, the overlapped eigenstates present avoid crossing with an energy gap $\Delta E \approx 0.5 J$.
	This fact means that the adiabatic condition can still be satisfied and that the topological pumping is immune to the disordered interaction in this experiment. 
	The numerical analysis here is in accordance with the experimental result in the main text, where the topological pumping is robust.
	On the other hand, we remark that slightly disordered interaction will suppress the spread of the wave packet during the pumping, which is beneficial. 

	\subsection{On-site energy offset}
\begin{figure}[!htbp]
		\centering
		\includegraphics[width=0.6\textwidth]{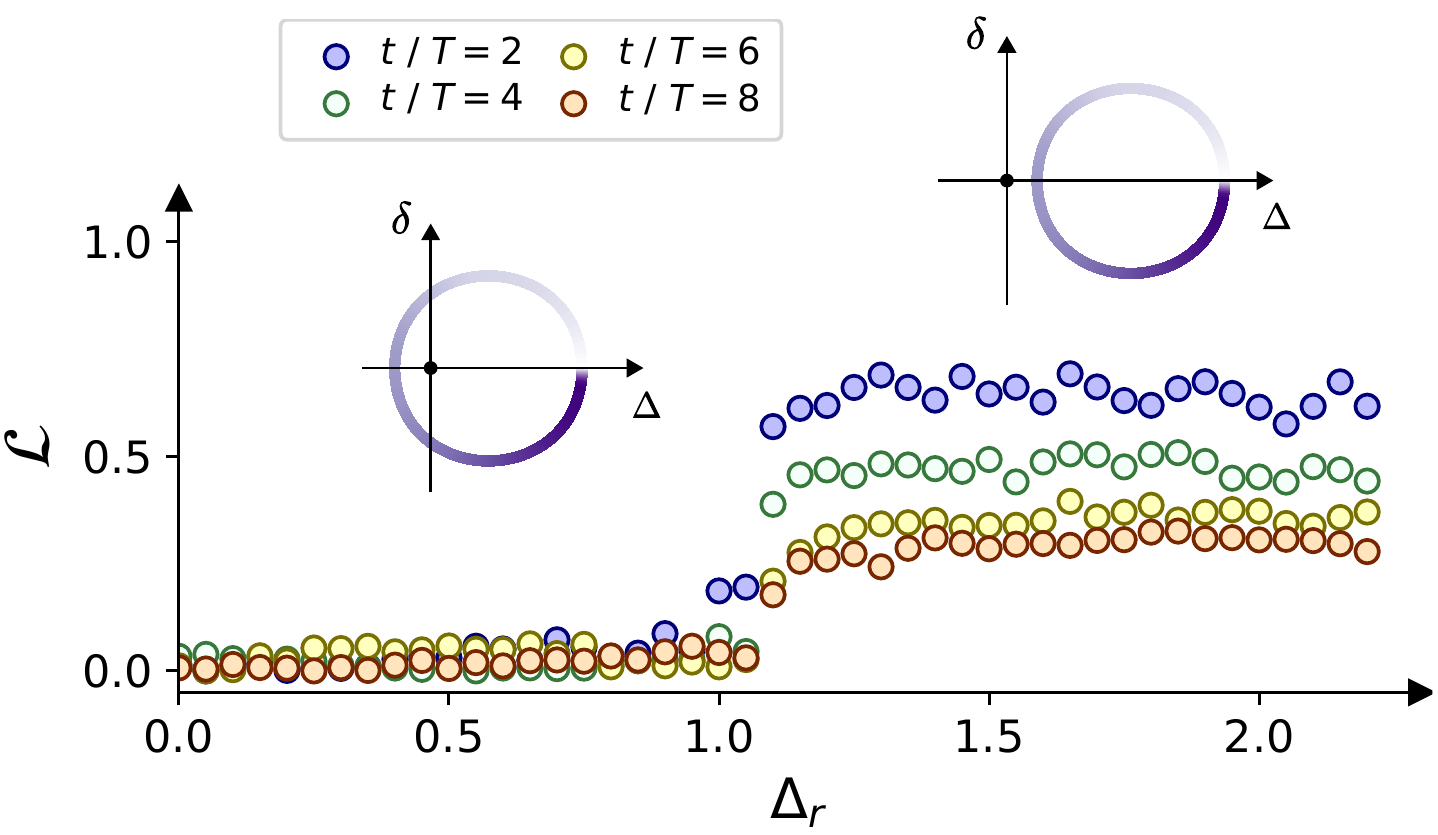}
		\caption{\label{supply_offset_scan} 
			Measured Loschmidt echo $\mathcal{L}$ against $\Delta_{r}$ under $t/T=2,4,6,8$.
		}
	\end{figure}

	In \rfig{supply_offset_scan}, we add a relative on-site energy offset $\Delta_{r}$ to $\Delta$, i.e., $\Delta = \Delta_0[\sin(\omega t)+\Delta_{r}]$, 
	and measure the Loschmidt echo $\mathcal{L}=\vert\langle \psi(t) \vert \psi(t=0)\rangle\vert^2$ which is the recovery of evolving quantum states~\cite{Xu2020,Zhao2022},where the initial state $\psi(t=0)=\vert 1_{j=18}\rangle$.
	When the relative offset $\Delta_r<1$, the degeneracy point remains encircled by the trajectory $\mathcal{C}$ in the $\Delta-\delta$ parameter plane, and the pump cycles are hardly affected by the offset, as indicated by  $\mathcal{L}\approx 0$.
	When $\Delta_r>1$, $\mathcal{C}$ no longer encircles the degeneracy point, as a consequence, the pumping cycles are dramatically changed, as indicated by the step change in $\mathcal{L}$.
	
	\begin{figure}[!htbp]
		\centering
		\includegraphics[width=0.6\textwidth]{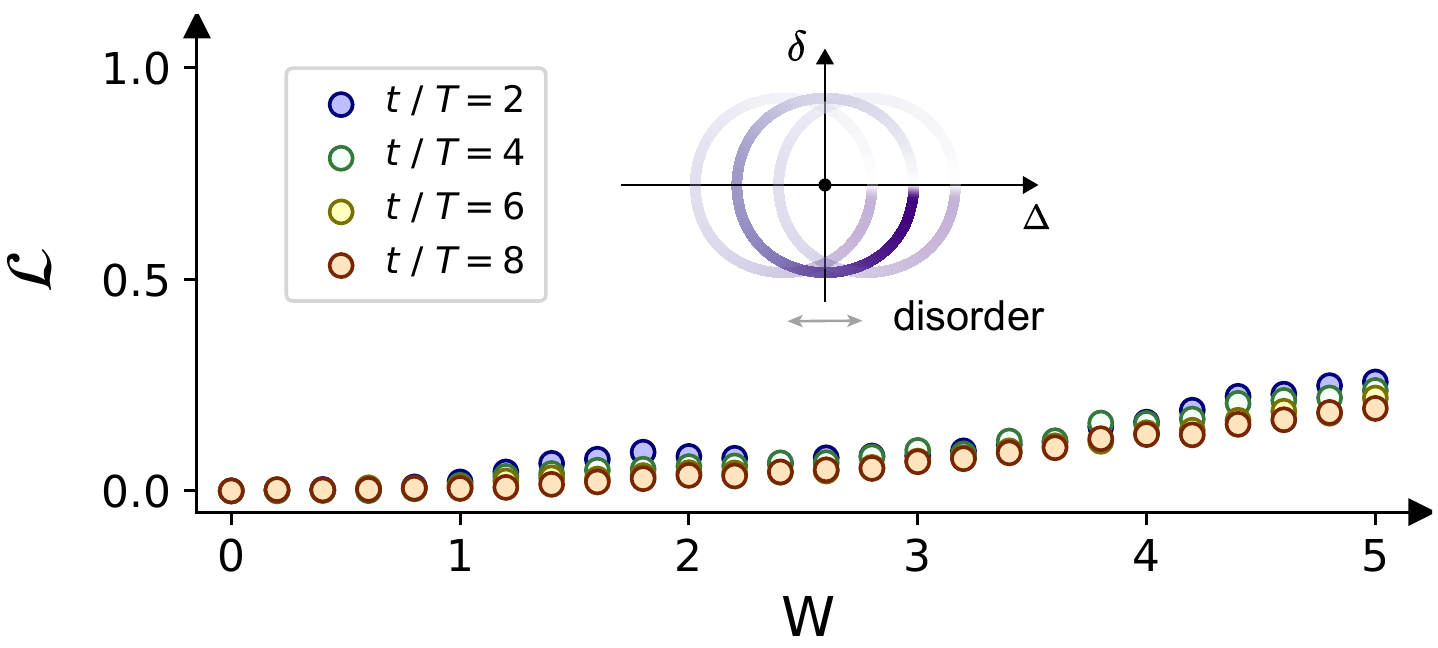}
		\caption{\label{supply_disorder_scan} 
			Measured Loschmidt echo $\mathcal{L}$ against $W$ under $t/T=2,4,6,8$, where each point is averaged over 50 times with random disorder to $\Delta$.
		}
	\end{figure}
	
	\subsection{On-site energy disorder}
\begin{figure}[!htbp]
		\centering
		\includegraphics[width=0.9\textwidth]{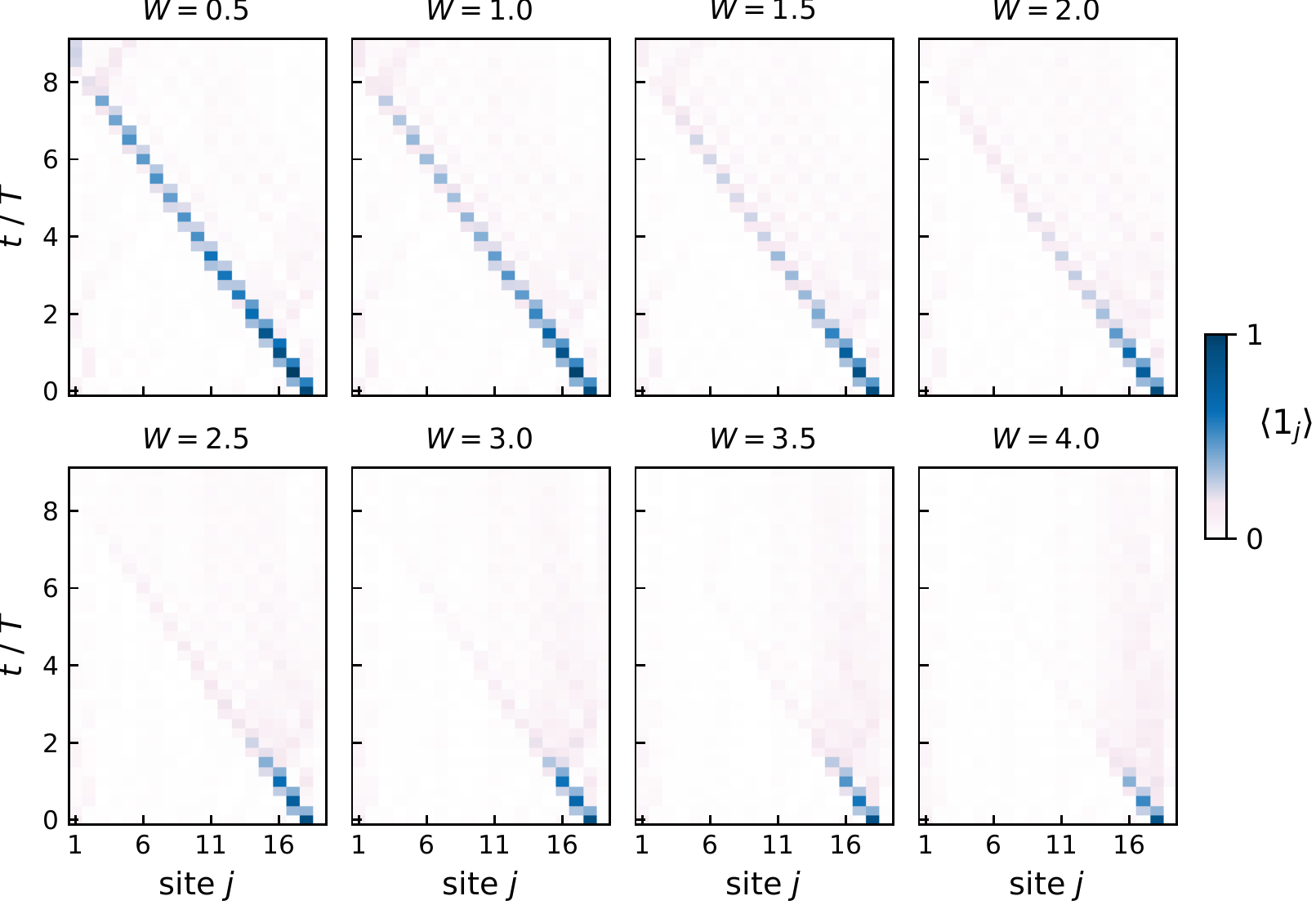}
		\caption{\label{supply_disorder_evo} 
			The population of $|n\rangle_j$ during the disordered pumping process with $W=$0.5, 1.0, 1.5, $\cdots$, 4.0.
		}
	\end{figure}

	In \rfig{supply_disorder_scan}, we add several random disorder to $\Delta$, with $\Delta = \Delta_0[\sin(\omega t)+ W \xi_l]$, where $\xi_l\in [-0.5,0.5]$ is a uniformly distributed random number, and $W$ is the disorder weight relative to $\Delta_0$.
	The evolution of particle populations  $|1_j\rangle$ under the on-site energy disorder is given in \rfig{supply_disorder_evo}, which shows transition from localized to delocalized population with increasing $W$.

	\clearpage

	
	%